\pdfoutput=1 
\documentclass[twocolumn]{aastex63}
\usepackage{longtable}
\usepackage{multirow}
\usepackage{booktabs}

\received{}
\revised{}
\accepted{}

\shorttitle{The progenitor system of Type Ia SN 2021hpr}
\shortauthors{Lim et al.}
\begin{document}
\title{THE EARLY LIGHT CURVE OF A TYPE Ia SUPERNOVA 2021hpr IN NGC 3147 : PROGENITOR CONSTRAINTS WITH THE COMPANION INTERACTION MODEL}

\correspondingauthor{Myungshin Im}
\email{myungshin.im@gmail.com}

\author[0000-0002-5760-8186]{Gu Lim}
\affiliation{SNU Astronomy Research Center, Seoul National University, 1 Gwanak-ro, Gwanak-gu, Seoul,  08826, Republic of Korea}
\affiliation{Astronomy Program, Department of Physics and Astronomy, Seoul National University, Gwanak-gu, Seoul 08826, Republic of Korea}
\affiliation{Department of Earth Sciences, Pusan National University, Busan 46241, Republic of Korea}

\author[0000-0002-8537-6714]{Myungshin Im}
\affiliation{SNU Astronomy Research Center, Seoul National University, 1 Gwanak-ro, Gwanak-gu, Seoul,  08826, Republic of Korea}
\affiliation{Astronomy Program, Department of Physics and Astronomy, Seoul National University, Gwanak-gu, Seoul 08826, Republic of Korea}

\author[0000-0002-6639-6533]{Gregory S. H. Paek}
\affiliation{SNU Astronomy Research Center, Seoul National University, 1 Gwanak-ro, Gwanak-gu, Seoul,  08826, Republic of Korea}
\affiliation{Astronomy Program, Department of Physics and Astronomy, Seoul National University, Gwanak-gu, Seoul 08826, Republic of Korea}

\author{Sung-Chul Yoon}
\affiliation{SNU Astronomy Research Center, Seoul National University, 1 Gwanak-ro, Gwanak-gu, Seoul,  08826, Republic of Korea}
\affiliation{Astronomy Program, Department of Physics and Astronomy, Seoul National University, Gwanak-gu, Seoul 08826, Republic of Korea}

\author{Changsu Choi}
\affiliation{SNU Astronomy Research Center, Seoul National University, 1 Gwanak-ro, Gwanak-gu, Seoul,  08826, Republic of Korea}
\affiliation{Astronomy Program, Department of Physics and Astronomy, Seoul National University, Gwanak-gu, Seoul 08826, Republic of Korea}
\affiliation{Korea Astronomy and Space Science Institute, 776 Daedeokdae-ro, Yuseong-gu, Daejeon 34055, Republic of Korea}

\author[0000-0002-0070-1582]{Sophia Kim}
\affiliation{SNU Astronomy Research Center, Seoul National University, 1 Gwanak-ro, Gwanak-gu, Seoul,  08826, Republic of Korea}
\affiliation{Astronomy Program, Department of Physics and Astronomy, Seoul National University, Gwanak-gu, Seoul 08826, Republic of Korea}

\author[0000-0003-1349-6538]{J. Craig Wheeler}
\affiliation{Department of Astronomy, University of Texas at Austin, Austin, TX, USA}
\author[0000-0002-0977-1974]{Benjamin P. Thomas}
\affiliation{Department of Astronomy, University of Texas at Austin, Austin, TX, USA}
\author[0000-0001-8764-7832]{Jozsef Vink{\'o}}
\affiliation{Department of Astronomy, University of Texas at Austin, Austin, TX, USA}
\affiliation{ Konkoly Observatory,  CSFK, Konkoly-Thege M. \'ut 15-17,
Budapest, 1121, Hungary}
\affiliation{ELTE E\"otv\"os Lor\'and University, Institute of Physics, P\'azm\'any P\'eter s\'et\'any 1/A, Budapest, 1117 Hungary}
\affiliation{Department of Optics \& Quantum Electronics, University of Szeged, D\'om t\'er 9, Szeged, 6720, Hungary}

\author[0000-0002-6925-4821]{Dohyeong Kim}
\affiliation{Department of Earth Sciences, Pusan National University, Busan 46241, Republic of Korea}

\author{Jinguk Seo}
\affiliation{SNU Astronomy Research Center, Seoul National University, 1 Gwanak-ro, Gwanak-gu, Seoul,  08826, Republic of Korea}
\affiliation{Astronomy Program, Department of Physics and Astronomy, Seoul National University, Gwanak-gu, Seoul 08826, Republic of Korea}

\author[0000-0003-3390-1924]{Wonseok Kang}
\affiliation{National Youth Space Center, Goheung, Jeollanam-do, 59567, Republic of Korea}
\author[0000-0003-4686-5109]{Taewoo Kim}
\affiliation{National Youth Space Center, Goheung, Jeollanam-do, 59567, Republic of Korea}

\author{Hyun-Il Sung}
\affiliation{Korea Astronomy and Space Science Institute, 776 Daedeokdae-ro, Yuseong-gu, Daejeon 34055, Republic of Korea}

\author{Yonggi Kim}
\affiliation{Chungbuk National University Observatory, 361-763 Cheongju, Republic of Korea}
\author{Joh-Na Yoon}
\affiliation{Chungbuk National University Observatory, 361-763 Cheongju, Republic of Korea}
\author{Haeun Kim}
\affiliation{Chungbuk National University Observatory, 361-763 Cheongju, Republic of Korea}
\author{Jeongmook Kim}
\affiliation{Chungbuk National University Observatory, 361-763 Cheongju, Republic of Korea}
\author{Hana Bae}
\affiliation{Chungbuk National University Observatory, 361-763 Cheongju, Republic of Korea}

\author[0000-0001-9730-3769]{Shuhrat Ehgamberdiev}
\affiliation{Ulugh Beg Astronomical Institute, Uzbek Academy of Sciences, 33 Astronomical Street, Tashkent 700052, Uzbekistan}
\affiliation{National University of Uzbekistan, Tashkent 100174, Uzbekistan}
\author[0000-0003-1169-6763]{Otabek Burhonov}
\affiliation{Ulugh Beg Astronomical Institute, Uzbek Academy of Sciences, 33 Astronomical Street, Tashkent 700052, Uzbekistan}
\author[0000-0003-0570-6531 ]{Davron Mirzaqulov}
\affiliation{Ulugh Beg Astronomical Institute, Uzbek Academy of Sciences, 33 Astronomical Street, Tashkent 700052, Uzbekistan}

%% https://en.wikibooks.org/wiki/LaTeX/Macros#New_commands

\begin{abstract} \label{abs:abstract}
    The progenitor system of Type Ia supernovae (SNe Ia) is expected to be a close binary system of a carbon/oxygen white dwarf (WD) and a non-degenerate star or another WD. Here, we present results from a high-cadence monitoring observation of SN 2021hpr in a spiral galaxy, NGC 3147, and constraints on the progenitor system based on its early multi-color light curve data. First, we classify SN 2021hpr as a normal SN Ia from its long-term photometric and spectroscopic data. More interestingly, we found a significant ``early excess'' in the light curve over a simple power-law $\sim t^{2}$ evolution. The early light curve evolves from blue to red and blue during the first week. To explain this, we fitted the early part of $BVRI$-band light curves with a two-component model of the ejecta-companion interaction and a simple power-law model. The early excess and its color can be explained by shock cooling emission due to a companion star having a radius of $8.84\pm0.58$\,$R_{\odot}$. We also examined HST pre-explosion images with no detection of a progenitor candidate, consistent with the above result. However, we could not detect signs of a significant amount of the stripped mass from a non-degenerate companion star ($\lesssim0.003\,M_{\odot}$ for H$\alpha$ emission). The early excess light in the multi-band light curve supports a non-degenerate companion in the progenitor system of SN 2021hpr. At the same time, the non-detection of emission lines opens a door for other methods to explain this event. 
\end{abstract}

\keywords{galaxies: distances and redshifts -- supernovae: general -- supernovae: individual (SN 2021hpr) -- methods: observational}


\section{Introduction} \label{sec:intro}
    The progenitor of Type Ia supernovae (SNe Ia) is expected to be a close binary system of a carbon/oxygen white dwarf (WD). There are two leading models for the progenitor systems of SNe Ia. One is the single degenerate (SD) model. In the SD model, SNe Ia explosion can result from thermonuclear runaway in the WD when the matter from its non-degenerate donors such as a main-sequence (MS), a subgiant (SG), a red giant (RG), or a helium star transfers its material over the Roche-lobe until the mass of WD approaches the Chandrasekhar-mass of $M_{\rm ch}\sim1.4\,M_{\odot}$ \citep{1973ApJ...186.1007W, 1982ApJ...253..798N, 1984ApJS...54..335I, 1996ApJ...470L..97H, 2014ApJ...794L..28W}. The other model, the double-degenerate (DD) model, predicts that a binary WD system can merge via emitting the gravitational wave radiation to produce an SN Ia \citep{1984ApJS...54..335I, 1984ApJ...277..355W}. SNe Ia have an empirical relation between the width of the light curve and the peak luminosity \citep[Width-Luminosity relation;][]{1993ApJ...413L.105P}. This allows SN Ia to be a good tool as a standardizable candle for measuring the distance and probing the expansion history of the universe \citep{1998AJ....116.1009R, 1999ApJ...517..565P}.

    Despite its usefulness for various astrophysical applications, SNe Ia progenitor systems are yet to be determined. One of the ways to constrain the progenitor system is to detect the shock-heated cooling emission (SHCE) in the early time light curve of SNe after the shock-breakout \citep{2010ApJ...708.1025K, 2011ApJ...728...63R, 2013ApJ...769...67P}. For the SD scenario, materials in the ejecta are heated by the shock which is produced from the collision between the ejecta and companion star \citep{2010ApJ...708.1025K}. While the ejecta expands and cools down, the emission (SHCE) can be detected at the ultraviolet(UV)/optical wavelengths. The brightness and duration of the SHCE depend on the radius of the companion and the viewing angle from the observer. For a $1\,M_{\odot}$ MS companion, the SHCE would be peaked at $B=-14$ AB magnitude while an RG companion ($\sim143\,R_{\odot}$) would produce it with $B\sim-18$ AB magnitude on a day after the explosion. Therefore, according to this picture, the luminosity of ``early excess'' on the rising part of the SNe Ia light curve can possibly constrain the companion star size (e.g., see \citealt{2019JKAS...52...11I}). The predicted SHCE is in general weak and lasts only for a few hours to days.

    There are other ways to produce this early excess without invoking the shock with the companion. If a sub-$M_{\rm ch}$ WD has a helium shell, a helium detonation on the shell can induce a shock wave traversing the carbon/oxygen core and trigger the second detonation in the CO core \citep[Double detonation or DDet;][]{1986ApJ...301..601W}. Sub-$M_{\rm ch}$ CO WD with a thin helium shell model ($<0.1\,M_{\odot}$) is favored in many recent simulations \citep{2007A&A...476.1133F, 2010A&A...514A..53F, 2013ApJ...770L...8P, 2019ApJ...873...84P} to explain a significant fraction of sub-luminous and normal SN Ia events. DDet models predict early excess due to high velocity nickel coming from the helium detonation \citep{2019ApJ...873...84P}. \citet{2019ApJ...873...84P} show that the color evolution during this early excess phase would have a red color peak (``red bump'' in \citet{2019ApJ...873...84P}).

    \citet{2016ApJ...826...96P} predicted the early excess of SNe Ia with various distributions of the radioactive nickel ($^{56}\rm Ni$) in the exploding WD and the presence of circumstellar material (CSM) around the primary WD. Shallow $^{56}\rm Ni$ distribution (Highly mixed) and extended CSM density can result in a bluer color evolution in the early phase. \citet{2020A&A...642A.189M} investigated that $^{56}\rm Ni$ shells in the outer ejecta can also produce the early excess in the light curve.

    A recent model suggests the early excess can also be seen in the DD system. \citet{2015MNRAS.447.2803L} predicted that a UV/Blue early emission can result from the interaction of SN ejecta and disk-originated matter (DOM), which forms an accretion disk surrounding the primary WD after the companion WD is tidally-disrupted.

    Observational studies show a diverse nature of SNe Ia early light curves. In some SNe Ia, no early excess is found, disfavoring the SD scenario. \citet{2011Natur.480..344N} constrained the progenitor radius of SN 2011fe as $<0.1\,R_{\odot}$ with the fireball model ($t^{2}$). \citet{2012ApJ...744L..17B} gave a tighter constraint on the companion size as $<0.02\,R_{\odot}$ favoring a DD system (but see \citealt{2014MNRAS.439.1959M} for a looser constraint). Some studies left a possibility of the SD model ruling out a red-giant companion for SN 2012ht \citep{2014ApJ...782L..35Y}, SN 2017cfd \citep{2020ApJ...892..142H}, and SN 2019ein where no early excess was reported (\citet{2020ApJ...893..143K} and Lim et al., in prep.).

    However, there are a number of studies showing a signature of early excess. \citet{2016ApJ...820...92M} constrained the companion mass of SN 2012cg to $6\,M_{\odot}$ MS binary companion using its early light curve and color evolution. But, other analyses of SN 2012cg favor a DD system for its progenitor \citep{2016MNRAS.459.1781L, 2018ApJ...855....6S, 2016MNRAS.457.3254M}. \citet{2017ApJ...845L..11H} detected a blue bump of SN 2017cbv inferring the presence of a subgiant companion of $56\,R_{\odot}$, although there is an aspect that the companion model does not fully describe the data. \citet{2015ApJS..221...22I} found a possible signal ($2\sigma$) of SHCE of SN 2015F. \citet{2019ApJ...872L...7L} suggested the DOM model can explain a linearly-rising flux at the early time of SN 2018oh better than the companion model. iPTF14atg \citep[][]{2015Natur.521..328C} and MUSSES1604D \citep[][]{2017Natur.550...80J} with the early UV flash and red bump are classified as a peculiar and a normal event with other peculiar features. SN 2018aoz \citep{2022NatAs.tmp...45N} and SN 2021aefx \citep{2022ApJ...932L...2A, 2022ApJ...933L..45H}, two normal SNe Ia, showed the early excess in their light curves, and various models are invoked to explain the early excess. Clearly, there is a need for more early light curve samples to better understand the SNe Ia progenitor system.

    Another way to constrain the progenitor model of SNe Ia is to detect spectral features resulting from the companion matter stripped/ablated by the ejecta \citep{1975ApJ...200..145W, 1986SvA....30..563C}. These features include hydrogen Balmer lines (e.g., H$\alpha$) and helium emission lines in the optical which are expected to be seen after the supernova enters the nebular phase \citep[$\gtrsim\,$200$\,$days;][]{2018ApJ...852L...6B}. \citet{2000ApJS..128..615M} predicted the mass of stripped matter with a numerical simulation as $0.15-0.17\,M_{\odot}$ for an MS ($820\,\rm km/s$), SG ($890\,\rm km/s$), and $0.53-0.54\,M_{\odot}$ for a RG companion star. \citet{2007PASJ...59..835M} also obtained a similar result. Many studies have investigated nebular spectra of SNe Ia, but estimating the mass of unbound material has been challenging \citep{2007ApJ...670.1275L, 2015A&A...577A..39L, 2019ApJ...872L..22T}. \citet{2016MNRAS.457.3254M} found a possible H$\alpha$ emission for SN 2013ct (corresponding to $\sim0.007\, M_{\odot}$) but no detection for the other 10 SNe Ia.

    As an attempt to enlarge a sample of SNe with early light curves, we have been carrying out the Intensive Monitoring Survey of Nearby Galaxies \citep[IMSNG;][]{2019JKAS...52...11I}. IMSNG monitors 60 nearby galaxies with a relatively high SNe rate at a cadence less than a day. Among the target galaxies is NGC 3147, where a Type Ia SN, SN 2021hpr, was discovered \citep{2021TNSTR.998....1I}. In this paper, we analyze the early, multi-band light curve of SN 2021hpr to constrain the explosion mechanism of SN 2021hpr with a focus given mainly on the SHCE model. Additionally, we report the result from a late phase spectroscopy of SN 2021hpr using the 9.2m Hobby-Eberly Telescope (HET) to find the Balmer emission lines that are expected to appear in the nebular phase for the SD model. All the magnitudes, other than quoted explicitly, are in the AB system.

    This paper is structured as the following. We present the observation, the data, and the data reduction procedures in Section 2. Section 3 shows the analysis results of the long-term and early light curves, spectral evolution, search for the SN progenitor in HST data, and deep spectroscopy for finding nebular emission lines. In Section 4, we discuss our main findings and demonstrate a wide variety of SNe Ia early color curves which may hint various explosion mechanisms for SNe Ia. Finally, we summarize our results in Section 5. We use $H_{\rm 0}$ value of $70\rm~km~s^{-1}~Mpc^{-1}$ \citep{2009ApJ...700..331H, 2016A&A...594A..13P}. We also neglect the $K-$ correction in our analysis since the redshift of NGC 3147 is very low at $z = 0.009346$ \citep[][]{2021TNSCR1031....1T}.

\section{Observation \& Data} \label{sec:observation}
    \subsection{Imaging Observation and Data}
        SN 2021hpr was discovered on 2021 April 2.45 UT \citep{2021TNSTR.998....1I}, and classified as a SN Ia \citep{2021TNSCR1031....1T}. Here, we report our IMSNG imaging observations and the data reduction procedures. In addition, we also used the data in the literature such as those taken at Caucasian Mountain Observatory \citep[CMO;][]{2021ATel14541....1T} and the Zwicky Transient Facility \citep[ZTF;][]{2019PASP..131a8002B}. We will describe how their data were transformed to our photometry system.

        Most of the data come from IMSNG. IMSNG monitoring observation provides the data to the 5$\sigma$ depths of $R \sim 19.5$ mag for a point source detection using a network of $0.4-1.0$ meter class telescopes around the world.

        NGC 3147 has been monitored by IMSNG since 2014 in $B$- and $R$-bands. In our data, SN 2021hpr was first identified in $B$- and $R$-band images taken on 2021 April 1.29 (UT) with the 1-m telescope of the Mt. Lemmon Optical Astronomy Observatory \citep[LOAO;][]{2019JKAS...52...11I}, located in the USA, after the last non-detection on 2021 March 31.18 with 3-$\sigma$ upper limits of $B=20.67$ mag and $R=20.18$ mag. Our first detection epoch precedes the discovery epoch of \citet{2021TNSTR.998....1I} by 1.1 days. The IMSNG data were taken nearly daily in the beginning, and then several times a day since the SN discovery up to +30 days from the $B$-maximum brightness of SN 2021hpr using $BVRI$-bands.

        In addition to the LOAO 1.0-m telescope, we used the 0.6-m telescope at Mt.Sobaek Optical Astronomy Observatory (SOAO), the 1.0-m telescope at Seoul National University Astronomical Observatory (SAO), the 0.6-m telescope at Chungbuk National University Observatory (CBNUO), the 1.0-m telescope at Deokheung Optical Astronomy Observatory (DOAO) in Korea, and SNUCAM \citep{2010JKAS...43...75I} of the 1.5-m telescope at Maidanak Astronomical Observatory in Uzbekistan \citep[MAO;][]{2018NatAs...2..349E}. Only $BVR$-band data were obtained at CBNUO and MAO. Refer to Table 2 in \citet{2019JKAS...52...11I} and Table 1 in \citet{2021JKAS...54...89I} for a detailed description of the facilities. For the SAO observation, we used the Finger Lake Instrumentation (FLI) KL4040 sCMOS camera. Each single exposure time varies with the observatory from 60 to 180 seconds.

        Standard reduction (bias, dark subtraction, and flat fielding) procedures were applied to the observed data using the \texttt{PyRAF} \citep{2012ascl.soft07011S} and the \texttt{Astropy} package \citep{Astropy13}. Additionally, we made a fringe pattern correction from the LOAO $I$-band data as described in \citep{2010ApJS..190..166J}. The astrometry calibration was conducted using \texttt{astrometry.net} \citep{2010AJ....139.1782L}.

        We performed photometry on images stacked from frames taken consecutively at a similar epoch (3 to 5 frames). The observation time of each combined image is defined as the median of the observing start times of each single frame used for stacking.

        We subtracted a reference image from science images using \texttt{HOTPANTS} \citep{2015ascl.soft04004B}, where reference images had been created in advance using images taken with the same telescope and instrument under the best observing conditions. Aperture photometry was performed on the subtracted images using \texttt{SExtractor} \citep{BA96} with an aperture diameter of 3$\times$Full Width Half Maximum (FWHM) of the Point Spread Function (PSF).

        The photometric calibration was conducted using stars from data release 1 (DR1) of Pan-STARRS\footnote{\url{https://catalogs.mast.stsci.edu/panstarrs/}} (PS1). The selection of the photometry reference stars and the calibration procedures are as follows.

        (i) Extended sources, QSO, variables, and transients were removed as flagged in the PS1 catalog within the field of view of each image. For CBNUO, we used sources around the image center within a radius of 75\% of the field of view to avoid systematic errors that may arise from image distortion around the edge.

        (ii) We further improved the point source selection by selecting sources with \texttt{i\_PSFmag}$-$\texttt{i\_Kronmag}$<0.05$\footnote{https://outerspace.stsci.edu/display/PANSTARRS/\\How+to+separate+stars+and+galaxies}. The PS1 magnitudes were transformed into the Johnson $BVRI$ system using equations in the form of $y=B_{0}+B_{1}x$, using coefficients $B_{0}$ and $B_{1}$ in Table 6 of \citet{2012ApJ...750...99T} as below.

        \begin{equation}
            (B-g_{PS1})=0.213+0.587(g-r)_{PS1}\,(\sigma_{B}=0.034),
        \end{equation}

        \begin{equation}
            (V-r_{PS1})=0.006+0.474(g-r)_{PS1}\,(\sigma_{V}=0.012),
        \end{equation}

        \begin{equation}
            (R-r_{PS1})=-0.138-0.131(g-r)_{PS1}\,(\sigma_{R}=0.015),
        \end{equation}

        \begin{equation}
            (I-i_{PS1})=-0.367-0.149(g-r)_{PS1}\,(\sigma_{I}=0.016).
        \end{equation}

        (iii) Next, we selected stars with the transformed magnitudes ranging\footnote{Bright sources ($< 13.5$ mag) in the PS1 catalog are known to be saturated \citep{2013ApJS..205...20M}. } from 13.5 to 17 with signal-to-noise larger than 10 and SExtractor \texttt{FLAG}=0. The magnitude zero points and their errors were taken as the mean and standard deviation of the zero points of reference stars. Typical zero-point errors are 0.005 to 0.185 magnitudes depending on filters and weather conditions.

        For the CMO data taken in the SDSS filter system, their $gri$-band magnitudes were transformed into the $BR$-band magnitudes using the equations in Table 2 of \citet{2007AJ....133..734B}.

        \begin{equation}
            B=g+0.2354+0.3915[(g-r)-0.6102]\,(\sigma_{g-r}=0.15),
        \end{equation}

        \begin{equation}
            V=g-0.3516-0.7585[(g-r)-0.6102]\,(\sigma_{g-r}=0.15),
        \end{equation}

        \begin{equation}
            R=r-0.0576-0.3718[(r-i)-0.2589]\,(\sigma_{r-i}=0.10),
        \end{equation}

        \begin{equation}
            I=i-0.0647-0.7177[(i-z)-0.2083]\,(\sigma_{i-z}=0.10).
        \end{equation}

        We used the $V$-band magnitude presented in \citet{2021ATel14541....1T}. For the ZTF photometry, we firstly converted $g_{ZTF}$-, and $r_{ZTF}$-band magnitudes into the PS1 filter system using the equations in \citet{2020RNAAS...4...38M}. These equations are expressed in terms of $g_{PS1}$, $r_{PS1}$ in this study.

        \begin{equation}
            g_{PS1}=0.948g_{ZTF}+0.052r_{ZTF}+0.011\,(\sigma_{g}=0.004),
        \end{equation}

        \begin{equation}
            r_{PS1}=0.076g_{ZTF}+0.924r_{ZTF}+0.004\,(\sigma_{r}=0.063).
        \end{equation}

        These ZTF-to-PS1 converted magnitudes were again transformed into the $BVR$-band magnitudes in the same way as above.

        We also cross-calibrated the photometry from different telescopes and references. We found that the magnitudes between LOAO and the other telescopes showed slight but significant systematic offsets (Table 1). The magnitude shifts were calculated by subtracting the LOAO magnitudes from the other telescopes' magnitudes in each band after the interpolation. Then, median values were added to the corresponding magnitudes to homogenize the magnitudes to the LOAO photometry. We did not calibrate the ZTF magnitudes since their photometric uncertainty is much larger than their magnitude difference.

        \begin{deluxetable}{ccccccc}\label{tab:table1}
            \tablecaption{Magnitude offsets ($\rm Mag_{LOAO}-Mag_{Telescopes}$) for other telescopes in each band.}
            \tablecolumns{7}
            \tablenum{1}
            \tablewidth{1pt}
            \tablehead{
            \colhead{} &
            \colhead{DOAO} &
            \colhead{SAO} &
            \colhead{CBNUO} & 
            \colhead{SOAO} & 
            \colhead{MAO} &
            \colhead{CMO} \\
            \colhead{} &
            \colhead{$(mag)$} &
            \colhead{$(mag)$} &
            \colhead{$(mag)$} &
            \colhead{$(mag)$} &
            \colhead{$(mag)$} &
            \colhead{$(mag)$}}
            \startdata 
                 $B$ & $-0.138$ & $-0.055$ & $-0.013$ & $-0.075$ & $-0.012$ & $0.069$ \\
                 $V$ & $-0.108$ & $-0.059$ & $-0.082$ & $-0.051$ & $-0.050$ & $-0.059$ \\
                 $R$ & $-0.127$ & $-0.107$ & $-0.103$ & $-0.047$ & $-0.072$ & $-0.106$ \\
                 $I$ & $-0.107$ & $-0.031$ &  &  &  & 0.184
            \enddata
        \end{deluxetable}

        \startlongtable
        \begin{deluxetable}{ccccccccccc}\label{tbl:table2}
        \tablecaption{Optical light curve of SN 2021hpr with no extinction corrected. The 3-$\sigma$ upper limits are also presented. The rest of this table is provided in an online machine-readable form. We also present the original CMO and ZTF data before the magnitude conversion.} 
        \tablecolumns{5}
        \tablenum{2}
        \tablewidth{0pt}
        \tablehead{\colhead{MJD} & \colhead{Phase} & \colhead{Magnitude} & \colhead{$\sigma_{\rm Mag}$} & \colhead{Telescope} &\\ \colhead{ } & \colhead{$(day)$} & \colhead{$(mag)$} & \colhead{$(mag)$} & }
        \startdata
        \multicolumn{5}{c}{B band}\\\hline
        $59256.85$ & $-65.01$ & $>21.12$ & - & $6$\\
        $59258.77$ & $-63.09$ & $>22.91$ & - & $6$\\
        $59259.83$ & $-62.02$ & $>23.07$ & - & $6$\\
        $59260.76$ & $-61.10$ & $>23.07$ & - & $6$\\
        $59261.71$ & $-60.15$ & $>23.00$ & - & $6$\\
        $59262.73$ & $-59.12$ & $>22.28$ & - & $6$\\
        $59275.26$ & $-46.59$ & $>19.89$ & - & $1$\\
        $59276.53$ & $-45.33$ & $>20.31$ & - & $1$\\
        $59277.26$ & $-44.60$ & $>21.62$ & - & $1$\\
        $59280.26$ & $-41.60$ & $>21.78$ & - & $1$\\
        $59283.33$ & $-38.53$ & $>21.89$ & - & $1$\\
        $59288.34$ & $-33.51$ & $>21.65$ & - & $1$\\
        $59295.12$ & $-26.74$ & $>20.78$ & - & $1$\\
        $59298.48$ & $-23.38$ & $>17.57$ & - & $5$\\
        $59298.48$ & $-23.38$ & $>17.69$ & - & $3$\\
        $59301.16$ & $-20.70$ & $>20.24$ & - & $1$\\
        $59302.21$ & $-19.64$ & $>19.45$ & - & $1$\\
        $59303.27$ & $-18.59$ & $>19.76$ & - & $1$\\
        $59303.54$ & $-18.31$ & $>18.09$ & - & $5$\\
        $59304.18$ & $-17.68$ & $>20.67$ & - & $1$\\
        $59304.92$ & $-16.94$ & $19.35$ & $0.06$ & $7$\\
        $59305.29$ & $-16.57$ & $18.72$ & $0.41$ & $8$\\
        $59305.29$ & $-16.56$ & $18.75$ & $0.09$ & $1$\\
        $59306.27$ & $-15.58$ & $18.02$ & $0.04$ & $1$\\
        $59307.16$ & $-14.69$ & $17.87$ & $0.20$ & $8$\\
        $59308.22$ & $-13.63$ & $17.22$ & $0.04$ & $1$\\
        $59308.38$ & $-13.47$ & $17.08$ & $0.04$ & $1$\\
        $59308.57$ & $-13.29$ & $17.11$ & $0.13$ & $2$\\
        $59308.65$ & $-13.21$ & $17.08$ & $0.11$ & $4$\\
        $59309.21$ & $-12.65$ & $16.50$ & $0.03$ & $1$\\
        $59309.34$ & $-12.52$ & $16.41$ & $0.04$ & $1$\\
        $59309.47$ & $-12.38$ & $16.49$ & $0.03$ & $3$\\
        $59309.54$ & $-12.32$ & $16.22$ & $0.07$ & $5$\\
        $59309.71$ & $-12.15$ & $16.09$ & $0.03$ & $7$\\
        $59310.19$ & $-11.66$ & $15.89$ & $0.04$ & $1$\\
        $59310.26$ & $-11.60$ & $15.86$ & $0.03$ & $1$\\
        $59310.34$ & $-11.52$ & $15.82$ & $0.04$ & $1$\\
        $59310.47$ & $-11.39$ & $15.87$ & $0.05$ & $2$\\
        $59310.63$ & $-11.23$ & $15.82$ & $0.01$ & $3$\\
        $59310.64$ & $-11.21$ & $15.80$ & $0.04$ & $4$\\
        $59311.19$ & $-10.67$ & $15.45$ & $0.04$ & $1$\\
        $59311.34$ & $-10.52$ & $15.41$ & $0.04$ & $1$\\
        $59311.63$ & $-10.23$ & $15.39$ & $0.02$ & $4$\\
        $59311.64$ & $-10.22$ & $15.29$ & $0.06$ & $5$\\
        $59311.82$ & $-10.04$ & $15.20$ & $0.03$ & $7$\\
        $59312.17$ & $-9.69$ & $15.12$ & $0.06$ & $1$\\
        $59312.35$ & $-9.50$ & $15.09$ & $0.05$ & $1$\\
        $59312.52$ & $-9.33$ & $15.16$ & $0.02$ & $3$\\
        $59312.72$ & $-9.13$ & $14.98$ & $0.01$ & $6$\\
        $59312.78$ & $-9.08$ & $14.90$ & $0.03$ & $7$\\
        $59313.23$ & $-8.63$ & $14.86$ & $0.04$ & $1$\\
        $59313.39$ & $-8.46$ & $14.83$ & $0.04$ & $1$\\
        $59313.57$ & $-8.29$ & $14.81$ & $0.04$ & $5$\\
        $59313.59$ & $-8.26$ & $14.94$ & $0.02$ & $3$\\
        $59313.72$ & $-8.14$ & $14.68$ & $0.03$ & $7$\\
        $59314.16$ & $-7.69$ & $14.68$ & $0.07$ & $1$\\
        $59314.37$ & $-7.48$ & $14.65$ & $0.04$ & $1$\\
        $59314.48$ & $-7.38$ & $14.80$ & $0.02$ & $3$\\
        $59314.65$ & $-7.21$ & $14.68$ & $0.03$ & $4$\\
        $59314.65$ & $-7.21$ & $14.74$ & $0.04$ & $3$\\
        $59315.22$ & $-6.64$ & $14.51$ & $0.04$ & $1$\\
        $59315.39$ & $-6.46$ & $14.53$ & $0.03$ & $1$%\hline
        \enddata
        \end{deluxetable}
        \tablecomments{In the column ``Telescope", LOAO, SAO, DOAO, SOAO, CBNUO, MAO, CMO, and ZTF corresponds to 1, 2, 3, 4, 5, 6, 7 and 8.}

    \begin{figure}[t]\label{fig:fig1}
        \centering
        \includegraphics[width=1\linewidth]{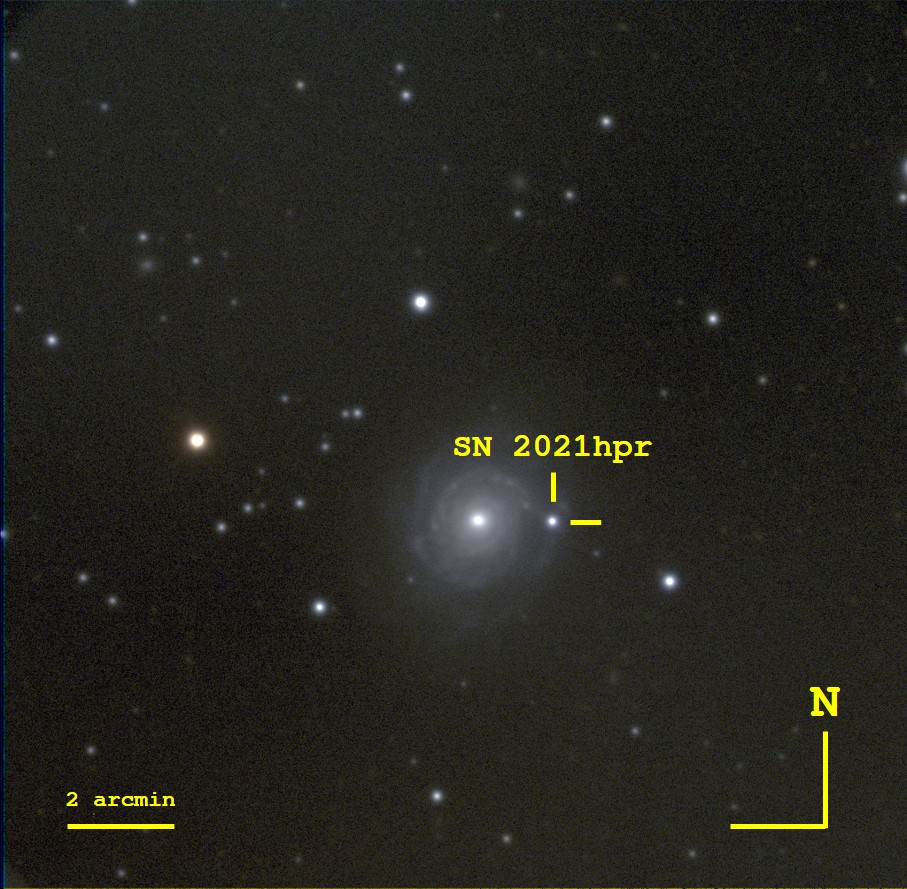}
        \caption{A color composite image of SN 2021hpr in NGC 3147. This image is composed of $R$-(red), $V$-(green), and $B$-(blue) band images observed on 2021-04-17 (almost at $B$-maximum) taken with the DOAO 1-m telescope. The yellow reticle points to the supernova. North is up, and East is to the left.}
    \end{figure}

    \subsection{SAO Spectroscopy}
        We performed long-slit spectroscopy on 2021 April 6, 14, and May 2 at SAO. We used the Shelyak LISA spectrograph\footnote{\url{https://www.shelyak.com/description-lisa/?lang=en}} with a grating of 300$\,\rm g/mm$ and a 2$\farcs$47 (50 $\mu$m) width slit. The slit angle was adjusted not to include the nucleus of NGC 3147. Bias, dark, and flat corrections by applying the standard IRAF procedures were conducted to the observed spectra. A $\rm Ne$ lamp was used for the wavelength calibration. Flux calibration was conducted using two standard stars, HR 4554 (A0V) and BD+75d 325 (O5P). The log of the SAO spectroscopy is provided in Table 3. The classification and spectral evolution of SN 2021hpr will be discussed in Section 3.5.

        \begin{deluxetable}{rrrr}\label{tab:table3}
            \tablecaption{SAO spectroscopic observation log for SN 2021hpr}
            \tablecolumns{4}
            \tablewidth{1.0\columnwidth}
            \tablenum{3}
            \tablewidth{1pt}
            \tablehead{
            \colhead{UT date} &
            \colhead{Phase\tablenotemark{$\rm \ast$} } &
            \colhead{Airmass} & 
            \colhead{Exposure} \\
            \colhead{ } &
            \colhead{(days)} &
            \colhead{ } &
            \colhead{(s)} 
            }
            \startdata 
                2021-04-06T14:37:02 & $-11.3$ & $1.28$ & $2\times900$ \\ 
                2021-04-14T13:55:36 & $-3.3$ & $1.27$ & $900$ \\
                2021-05-02T13:09:12 & $+14.7$ & $1.29$ & $8\times900$ \\
            \enddata 
            \tablenotetext{\ast}{Relative to the epoch of $B$-band maximum ($59321.864\,\rm MJD$)}
        \end{deluxetable}  

\section{Results} \label{sec:analysis}
    \subsection{Optical long-term light curve} 
        Table 2 provides the optical light curve data before the dust extinction correction, and Figure 3 shows the optical light curve from $-17.025$ to $29.329$ days from the $B$-maximum, corrected for the galactic and host extinction. The galactic reddening is adopted as $E(B-V)_{\rm MW}=0.021$ \citep[$A_{B}=0.088$, $A_{V}=0.067$, $A_{R}=0.053$, and $A_{I}=0.037$;][]{2011ApJ...737..103S}. The host galaxy reddening is determined from the peak $B-V$ color as described in Section 3.2, and we assumed the galactic extinction curve \citep[][]{1999PASP..111...63F} and $R_{V}=3.1$ to obtain the extinction correction in each band. The light curves of SN 2011fe, one of the most well-studied SNe Ia, are over-plotted for comparison, matching the $B$-band peak brightness epoch and giving arbitrary y-direction shifts to overlap with the maximum brightness of SN 2021hpr. For this, we adopt the SN 2011fe's peak time as 55814.48 MJD \citep{2016ApJ...820...67Z}. A polynomial fit (solid line) was performed on the SN 2021hpr light curve using the data near the peak from $-7$ to $+18$ days, which gives us the times and brightnesses ($t_{\rm \lambda, max}$, $m_{\rm \lambda, max}$) at the peak brightness and the decline rates, $\Delta m_{\rm 15}(\lambda)$, at different bands. These quantities are taken from the 50 percentile value in the distribution using the bootstrap re-sampling of the light curves ($N=1000$; Table 4). The uncertainty is adopted from the standard deviation in the distribution of each parameter.

    \begin{deluxetable*}{cccccc}\label{tab:table4}
        \tablecaption{Light curve parameters, estimated from the polynomial fit of the light curve of SN 2021hpr. Both the galactic and host extinctions are corrected. Also included are the peak absolute magnitudes ($M_{\rm \lambda, max}$) from \citet{2022PASP..134g4201Z}. All magnitudes in this table are in the AB system.}
        \tablecolumns{5}
        \tablenum{4}
        \tablewidth{1pt}
        \tablehead{
        \colhead{} &
        \colhead{$t_{\rm \lambda, max}$} &
        \colhead{$m_{\rm \lambda, max}$} &
        \colhead{$\Delta m_{15}(\lambda)$} & 
        \colhead{$M_{\rm \lambda, max}$} &
        \colhead{$M_{\rm \lambda, max}$ (Zhang+22)} \\
        \colhead{} &
        \colhead{$\rm (MJD)$} &
        \colhead{$(mag)$} &
        \colhead{$(mag)$} &
        \colhead{$(mag)$} &
        \colhead{$(mag)$}
        }
        \startdata 
             $B$ & $59321.856\pm0.218$ & $13.724\pm0.004$ & $0.988\pm0.026$ & $-19.553\pm0.111$ & $-19.621\pm0.210$ \\ 
             $V$ & $59324.093\pm0.127$ & $13.811\pm0.003$ & $0.715\pm0.022$ & $-19.226\pm0.111$ & $-19.349\pm0.210$ \\
             $R$ & $59323.456\pm0.124$ & $13.973\pm0.008$ & $0.721\pm0.031$ & $-19.109\pm0.111$ & $-19.127\pm0.210$ \\
             $I$ & $59320.697\pm0.137$ & $14.762\pm0.004$ & $0.486\pm0.008$ & $-18.400\pm0.111$ & $-18.595\pm0.210$ \\
        \enddata 
    \end{deluxetable*}

    \subsection{The reddening, peak absolute magnitude, and distance to NGC 3147} 
        We measured the host reddening using the relation between the intrinsic $B-V$ color at the maximum brightness, $(B_{\rm max}-V_{\rm max})_{0}$, and $\Delta m_{\rm 15}(B)$ \citep{1999AJ....118.1766P}. The observed $B-V$ color at the maximum brightness, $(B_{\rm max}-V_{\rm max})_{\rm corr}$, is $-0.004\pm0.005$ after the galactic extinction correction alone. The intrinsic color in maximum brightness, $(B_{\rm max}-V_{\rm max})_{0}$, is expected to be $-0.083\pm0.040$ for $\Delta m_{15}(B)$ = $0.988\pm0.026$. From these values, we measure $E(B-V)_{\rm host}$ as $0.079\pm0.040$. Therefore, the sum of the MW and host color excess, $E(B-V)_{\rm total}$, is $0.100$ $(0.021+0.079)$\,mag. Note that the reddening of the host is smaller than $E(B-V)_{\rm host} = 0.22\pm0.05$\,mag for SN 2008fv \citep{2012A&A...537A..57B}, an SN Ia that appeared in another arm of NGC 3147. The extinction values in each band are estimated assuming the galactic extinction curve \citep[$R_{V}=3.1$;][]{1999PASP..111...63F}.

        In Figure 2, we show previous distance estimates that are derived from historical SNe Ia \citep{2010ApJ...716..712A, 2006ApJ...647..501P, 2013AJ....146...86T, 2006ApJ...645..488W, 2008MNRAS.389.1577T, 2012A&A...537A..57B, 2008ApJ...686..749K, 2007ApJ...659..122J, 2005ApJ...624..532R, 2000ApJ...540..634P} and the Tully-Fisher (TF) relation (\citealt{1984A&AS...56..381B, 1986A&A...156..157B, 1988cng..book.....T}). Note that the distances from TF are measured before 1990. These values range from 30 to 50 Mpc with a median value of $41.7\,\rm Mpc$, and converted to appropriate values adopting a common Hubble constant of $70\rm~km~s^{-1}~Mpc^{-1}$.

        Considering the historical distances ($41.7\,\rm Mpc$), we obtained $M_{B, \rm max}$ as $-19.40\pm0.20$. Figure 4 shows the width-luminosity relation including SNe Ia from the CfA3 catalog \citep[gray dots;][]{2009ApJ...700..331H}, SN 2021hpr (the yellow star), and SN 2011fe \citep[the blue circle;][]{2016ApJ...820...67Z}. We plot 105 SNe Ia from the CfA3 sample \citep{2009ApJ...700..331H} with both the photometry and the decline rates available for $B$-band in the AB system. Since the position of SN 2021hpr is near the Phillips relation, we deduce that SN 2021hpr is a normal SN Ia photometrically. Adopting the recent Cepheid distance value from \citet{2022arXiv220910558W} of $40.1$ Mpc ($\mu$ = 33.014\,mag) brings the $M_{B, \rm max}$ value only by 0.1\,mag to a fainter side, and does not affect this conclusion.

        Alternatively, we estimate $M_{B, \rm max}$ as $-19.55\pm0.11$ AB mag using the decline rate ($\Delta m_{15}(B)$ = $0.988\pm0.026$) from the Phillips relation (Figure 4), which is consistent with a recent measurement in \citet{2022PASP..134g4201Z}. We find that the distance modulus ($\mu$) of NGC 3147 is 33.28$\pm$0.11\,mag or the distance ($d$) as $45.23\pm2.31\,\rm Mpc$. Our value is a bit larger than the median value of the distribution in Figure 2. However, our value agrees well with 43.70 Mpc derived from a modern day SN Ia, SN 2008fv \citep{2012A&A...537A..57B}. In addition, we provide the peak absolute magnitudes from \citet{2022PASP..134g4201Z} in the AB system using their distance modulus of $33.46\pm0.21\rm \,mag$ to compare our measurements with them (Table 4). Our measurements also agree with the value from \citet{2022PASP..134g4201Z} within their error. For further analysis, we determined to use the distance modulus of $\mu = 33.28\pm0.11$\, mag.

    \begin{figure}[t]
        \centering
        \includegraphics[width=1.0\linewidth]{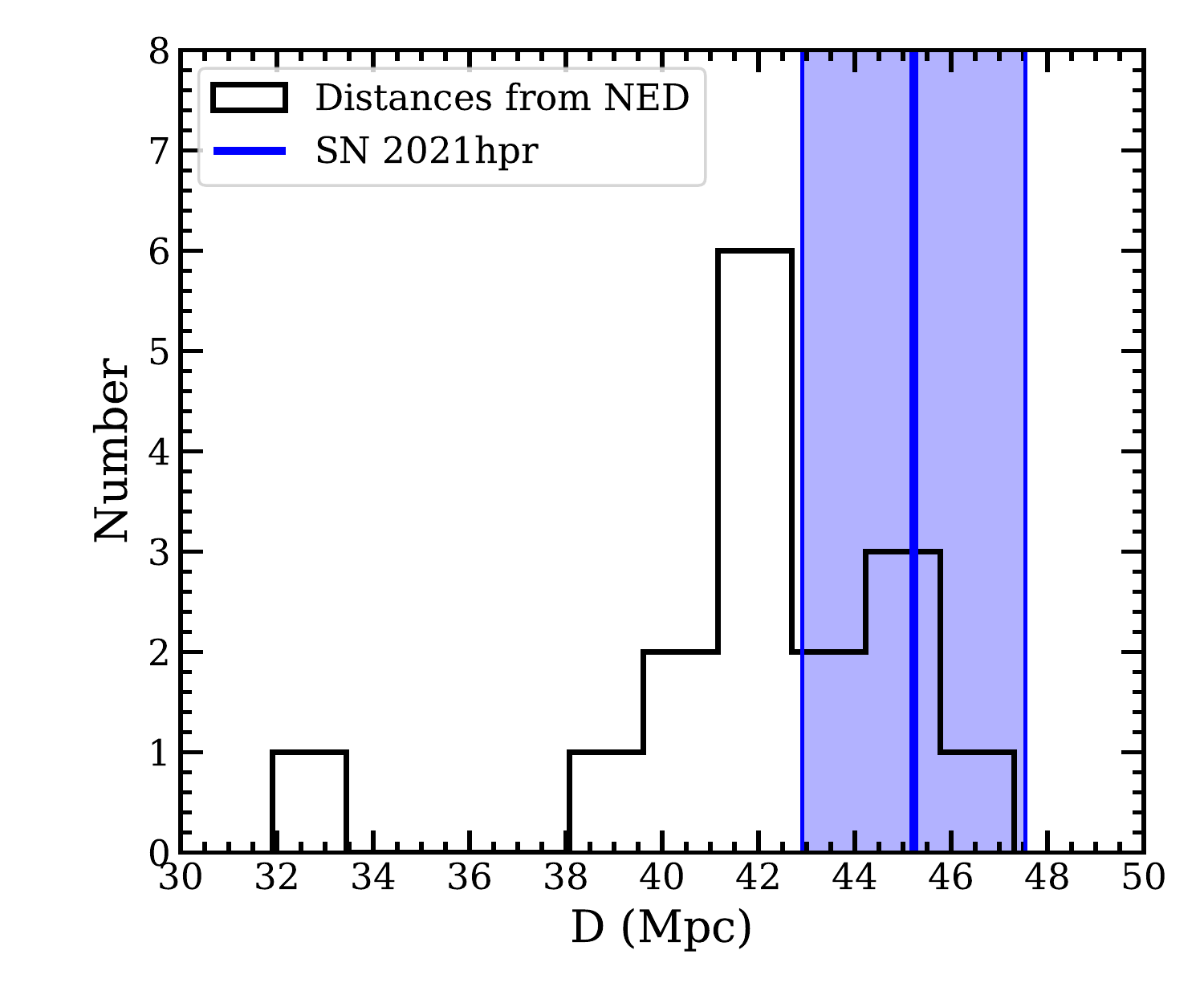}
        \caption{A histogram of estimated distances of NGC 3147 from the NED database. The distance estimated in this study ($45.23\pm2.31\,\rm Mpc$) is marked as the blue solid line shaded with a $1\sigma$ uncertainty.}
    \end{figure}

    \begin{figure*}[t]
        \centering
        \includegraphics[width=1.0\textwidth]{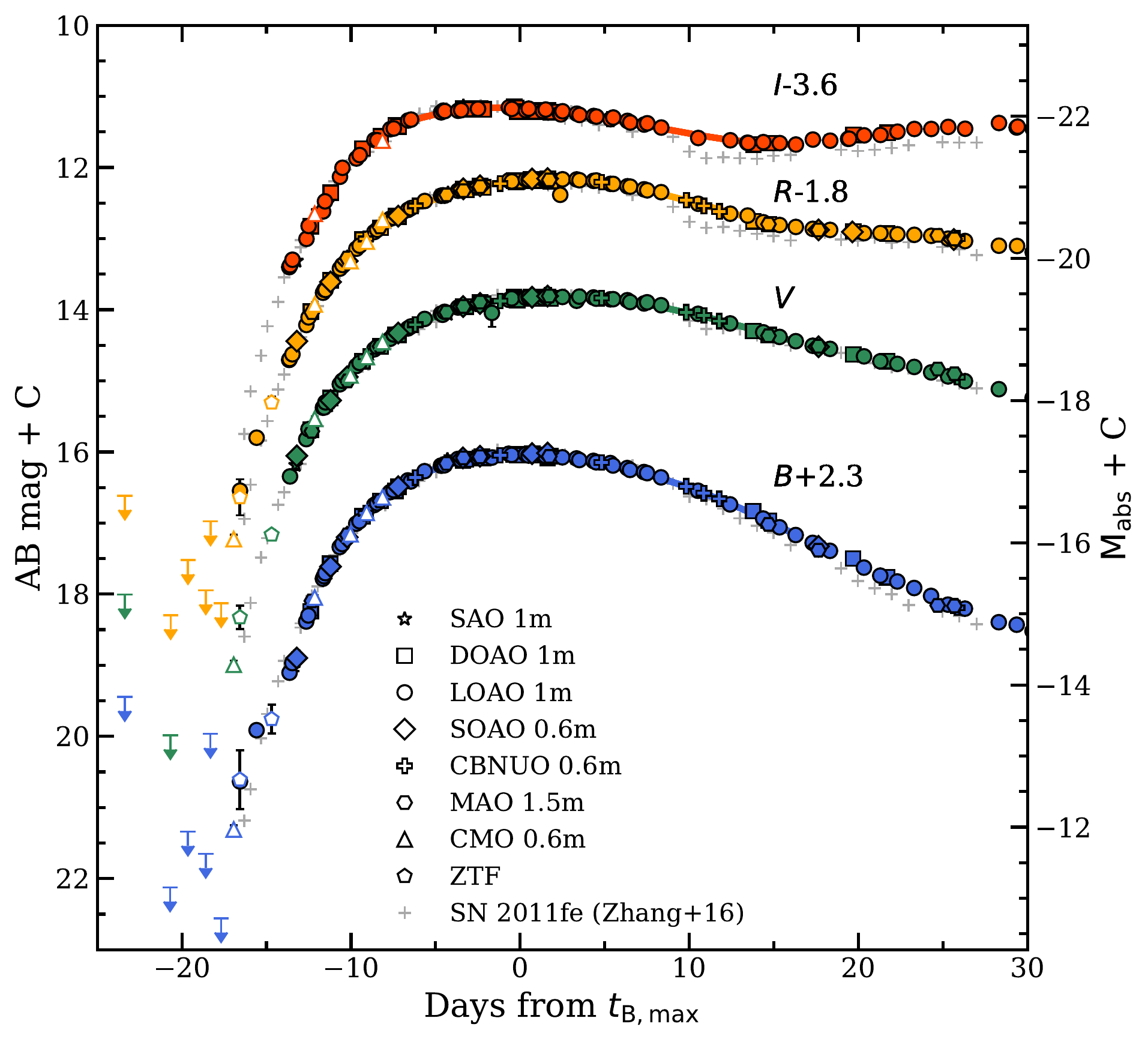}
        \caption{The optical light curve of SN 2021hpr. Other reported data points like CMO and ZTF are marked as open symbols. Arrows are 3$\sigma$ upper limits. The polynomial fitting results are over-plotted as solid line. SN 2011fe is also marked as gray cross symbols with an offset to the y-axis direction. The milky way and the host galaxy extinction are both corrected.}
    \end{figure*}

    \begin{figure}[t]
        \centering
        \includegraphics[width=1.0\linewidth]{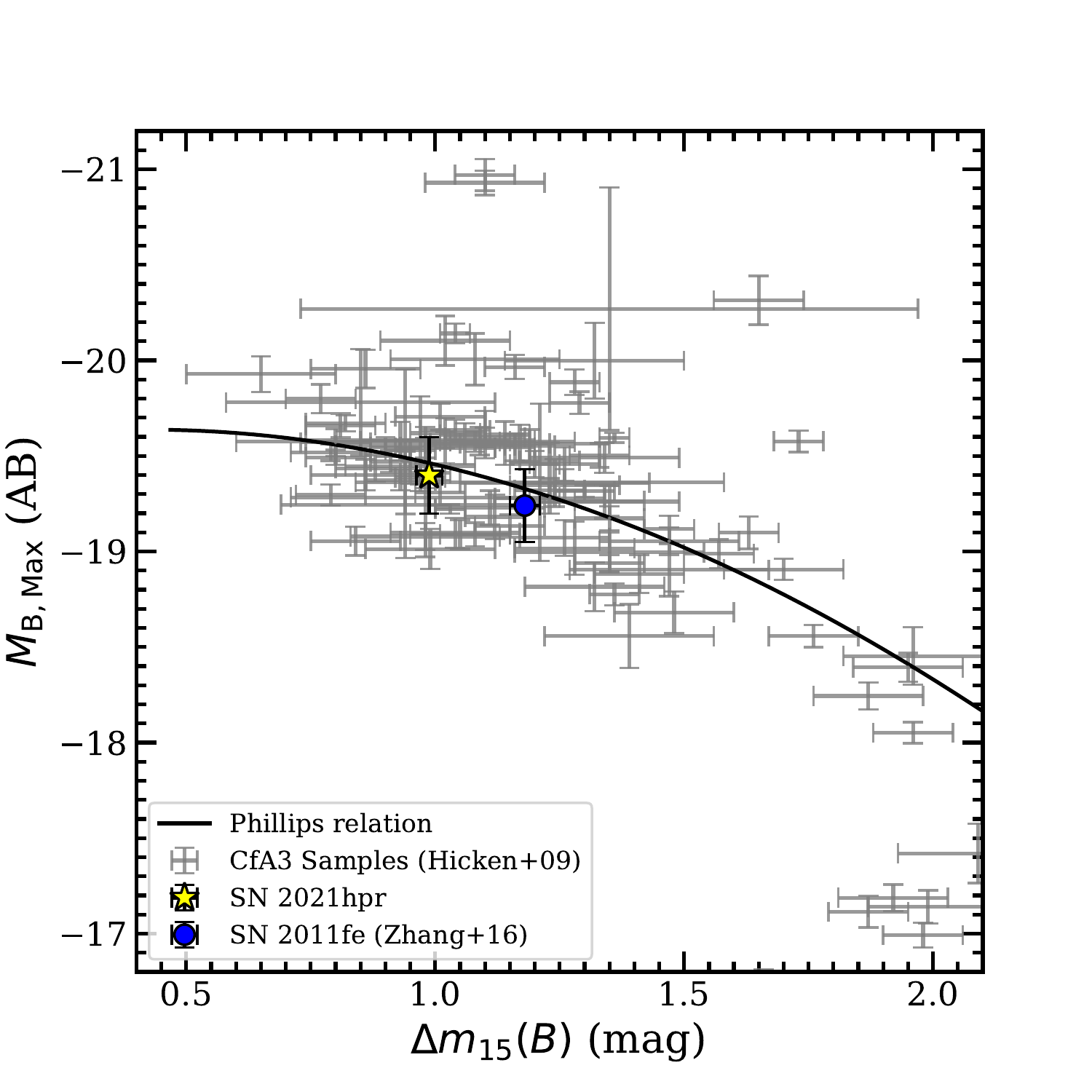}
        \caption{The location of SN 2021hpr (a yellow star) in the width-luminosity relation. The Phillips relation is marked in a black solid line. For the $M_{B, \rm max}$ value and its error of SN 2021hpr in this plot, we assumed the median distance and its standard deviation from the literature in Figure 3. SNe Ia from the CfA3 catalog are plotted in gray dots with error bars. SN 2011fe are plotted with a blue circle.}
    \end{figure}

    \subsection{The early light curve}
        Figure 5 shows the early light curve of SN 2021hpr $\pm5$ days from the first light time ($t_{\rm fl}$\footnote{We adopted the term ``the first light time" from \citet{2015ApJS..221...22I} describing that the photons generated from radioactive decay of $^{56}\rm Ni$ start to escape from the photosphere. The first light time is distinguished from ``the explosion time'' which is the actual explosion time.}). In general, early flux evolution of SNe Ia can be described well with a rising power law ($t^{\alpha}$) with $\alpha \sim 2$ \citep[the fireball model;][]{1998AJ....116.1009R, 2011Natur.480..344N}, However, SN 2021hpr shows a bumpy feature in the very early time ($t_{\rm fl} < 0$) that seems to deviate from a simple power-law light curve. Here, we examine this early excess in the light curve quantitatively using a power-law model and an ejecta-companion interaction model suggested by \citet{2010ApJ...708.1025K}. We modeled the rising part of the SN Ia light curve with a combination of a simple power-law and a SHCE. The simple power-law model is described in Equation (11).

        \begin{equation}
            M(t) = M_{0}-2.5 \alpha \log_{10}(t-t_{\rm fl}).
        \end{equation}

        Here, $M(t)$ is the absolute magnitude as a function of time $t$, and $M_{0}$ is a normalization factor of the absolute magnitude at a unit time of $t-t_{\rm fl} = 1$\,day.

        The SHCE light curve is calculated using the ejecta-companion interaction model of \citet[][hereafter K10]{2010ApJ...708.1025K}. To calculate the effective temperature, $T_{\rm eff}(t)$, and the luminosity, $L(t)$, of SHCE, we use the equations below which are taken from \citet{2015ApJS..221...22I}.

        \begin{equation}
            T_{\rm eff}(t) = 5.3\times10^{3} \frac{R^{1/4}_{10}}{\kappa^{35/36}_{0.2}} t^{-37/72}_{\rm exp}~{\rm K},
        \end{equation}

        \begin{equation}
            L(t) = 2.0\times10^{40} \frac{R_{10}M^{1/4}_{c}v_{9}^{7/4}}{\kappa_{0.2}^{3/4}} t_{\rm exp}^{-0.5}~{\rm erg~s^{-1}}.
        \end{equation}

        Here, $R_{10}$ is the radius of the companion star in units of $10^{10}\rm~cm$ ($R_{10}=R_{*}/10^{10}$ and $R_{*}=a/2$ where $a$ and $R_{*}$ are separation distance and stellar radius), $\kappa_{0.2}$ is the opacity in units of $0.2\rm~cm^{2}~g^{-1}$ which is adopted as $1.0$, $M_{c}$ is the ejecta mass in units of $1.4\,M_{\odot}$  which is adopted as $1/1.4$, $t_{\rm exp}$ is the time since the explosion in units of days, and $v_{9}$ is the expansion velocity of the ejecta in units of $10^{9} \rm~cm~s^{-1}$ (adopted as $1.0$, here).

        The light curve fit was performed on the 4 band data simultaneously by minimizing chi-square using the \texttt{Python} library of \texttt{LMFIT} \citep{2014zndo.....11813N}. The fitting was performed using the data points from MJD=59304.92 to 59311.35, i.e., all the detection points presented in Figure 5. Free parameters are $\alpha_{\rm B}$, $\alpha_{\rm V}$, $\alpha_{\rm R}$, $\alpha_{\rm I}$, $M_{\rm 0,B}$, $M_{\rm 0,V}$, $M_{\rm 0,R}$, $M_{\rm 0,I}$, $t_{\rm fl}$, $R_{*}$, and $t_{\rm gap}$, where the last two parameters are the radius of the companion star in units of $R_{\odot}$ and the time gap between $t_{\rm exp}$ and $t_{\rm fl}$ ($t_{\rm gap}$ = $t_{\rm fl} - t_{\rm exp}$), where $t_{\rm exp}$ is the explosion time which marks the start of SHCE. Furthermore, the amount of SHCE in the companion model is dependent on the viewing angle. We assume the optimal viewing angle that the observer looks at the shocked region on the line of sight (Observer--Companion--WD), giving the brightest collision luminosity. If we assume a common viewing angle, the companion radius could be larger than that obtained with the optimal viewing angle by about a factor of 10.

        The best-fit parameters and the best-fit light curves are given in Table 5 and Figure 5. As seen in Figure 5, our two-component model can explain the early excess of SN 2021hpr with a $R_{*}=8.84\pm0.58\,R_{\odot}$ sized companion, giving the goodness of fit of $\chi^{2}_{\nu}$ as $2.4$. The explosion time ($t_{\rm exp}$) is estimated as $59304.73\pm0.01$ MJD. $t_{\rm gap}$ is also estimated as $1.81\pm0.18\,$days. This can be regarded as a large value (e.g., \citealt{2017MNRAS.472.2787N}), but it is acceptable if a large fraction of $^{56}\rm Ni$ is deep from the ejecta surface in which case $t_{\rm gap}$ can be up to a few days after the explosion \citep{2013ApJ...769...67P}. When assuming a common viewing angle which we modeled by multiplying Eq. (13) by 0.1, the radius of the companion star can be $R_{*}=175.59\pm52.70\,R_{\odot}$ ($\chi^{2}_{\nu}=3.1$, see the top left panel of Figure A.1). When we fix the power index to $2$ (the fireball model), $t_{\rm gap}$ is $0.91\pm0.13\,$days and $R_{\rm *}=6.52\pm1.09\,R_{\odot}$ with $\chi^{2}_{\nu}=5.3$ (Optimal viewing angle). A simple power-law model gives a poorer fit ($\chi^{2}_{\nu}=5.0$, dash-dotted line) and a too large $\alpha$. Furthermore, the best-fit $t_{\rm fl}$ from the simple power-law model is several days before the best-fit $t_{\rm fl}$ from the two component model and the best-fit model curve goes over the 3-$\sigma$ detection limits at the last non-detection. Note that the best-fit $t_{\rm fl}$ from this model is $59301.10\pm0.52\,\rm MJD$). Therefore, a simple power-law model is disfavored.

        \begin{deluxetable*}{ccccccccc}[t]\label{tab:table5}
            \tablecaption{The best results of the early light curve fit by different methods. We did not include the case of the Companion+Simple power-law model (fireball) assuming the common viewing angle because we obtained unacceptable result that model lines cannot explain the upper limit in observed data.}
            \tablecolumns{9}
            \tablenum{5}
            \tablehead{
                \colhead{Fitting method} &
                \colhead{Viewing angle} &
                \colhead{$\alpha$} &
                \colhead{$M_{0}$} &
                \colhead{$t_{\rm exp}$} &
                \colhead{$t_{\rm fl}$} &
                \colhead{$t_{\rm gap}$} &
                \colhead{$R_{*}$} &
                \colhead{$\chi^{2}_{\nu}$}\\ 
                \colhead{} &
                \colhead{} &
                \colhead{} &
                \colhead{($mag$)} &
                \colhead{(MJD)} &
                \colhead{(MJD)} &
                \colhead{($days$)} &
                \colhead{($R_{\odot}$)} &
                \colhead{}
            }
            \startdata 
                \multirow{4}{*}{Simple power-law} & \multirow{4}{*}{ } & ($B$) $3.84\pm0.32$ & $24.70\pm0.99$ & & & & &\multirow{4}{*}{$5.0$} \\ 
                 & & ($V$) $3.56\pm0.27$ & $23.95\pm0.86$ & & $59301.10$ & & \\
                 & & ($R$) $3.45\pm0.28$ & $23.84\pm0.88$ & & $\pm0.52$ & &  \\
                 & & ($I$) $3.44\pm0.27$ & $24.38\pm0.83$ & & & &  \\ \hline
                \multirow{8}{*}{Companion+Simple power-law} & \multirow{4}{*}{Optimal} & ($B$) $2.03\pm0.13$  & $18.44\pm0.29$ &  &  & & & \multirow{4}{*}{$2.4$}\\ 
                 & &($V$) $1.48\pm0.10$ & $17.59\pm0.22$ & $59304.73$& $59306.54$ & $1.81$ & $8.84$ &  \\
                 & &($R$) $1.50\pm0.11$ & $17.83\pm0.23$ & $\pm0.01$ & $\pm0.18$ & $\pm0.18$ & $\pm0.58$ & \\
                 & &($I$) $1.79\pm0.13$ & $18.99\pm0.26$ & & & & \\  
                  \cmidrule{2-9} & \multirow{4}{*}{Common} & ($B$) $3.74\pm0.28$ & $23.37\pm0.78$ & & & & & \multirow{4}{*}{$3.1$} \\
                 & &($V$) $2.74\pm0.18$ & $21.17\pm0.53$ & $59304.30$ & $59303.55$ & $-0.75$& $175.59$\\
                 & &($R$) $2.58\pm0.19$ & $21.02\pm0.55$ & $\pm0.11$ & $\pm0.45$ & $\pm0.43$& $\pm52.71$\\
                 & &($I$) $2.79\pm0.20$ & $22.05\pm0.56$ & & & & \\ \hline
                \multirow{3}{*}{Companion+Simple power-law} & \multirow{4}{*}{Optimal} & \multirow{4}{*}{$2$ (Fixed)} & ($B$) $18.85\pm0.06$ & & & & & \multirow{4}{*}{$5.3$} \\ 
                 & & & ($V$) $18.79\pm0.06$ & $59304.70$ & $59305.60$ & $0.91$ & $6.52$\\
                (Fireball) & & & ($R$) $19.01\pm0.05$ & $\pm0.03$ & $\pm0.13$ & $\pm0.13$ & $\pm1.09$\\
                 & & & ($I$) $19.70\pm0.05$ & & & &
            \enddata 
        \end{deluxetable*}

        \begin{figure*}\label{fig:fig5}
        \centering
        \begin{minipage}[c]{0.8\linewidth}
        \centering
                \includegraphics[width=0.8\textwidth]{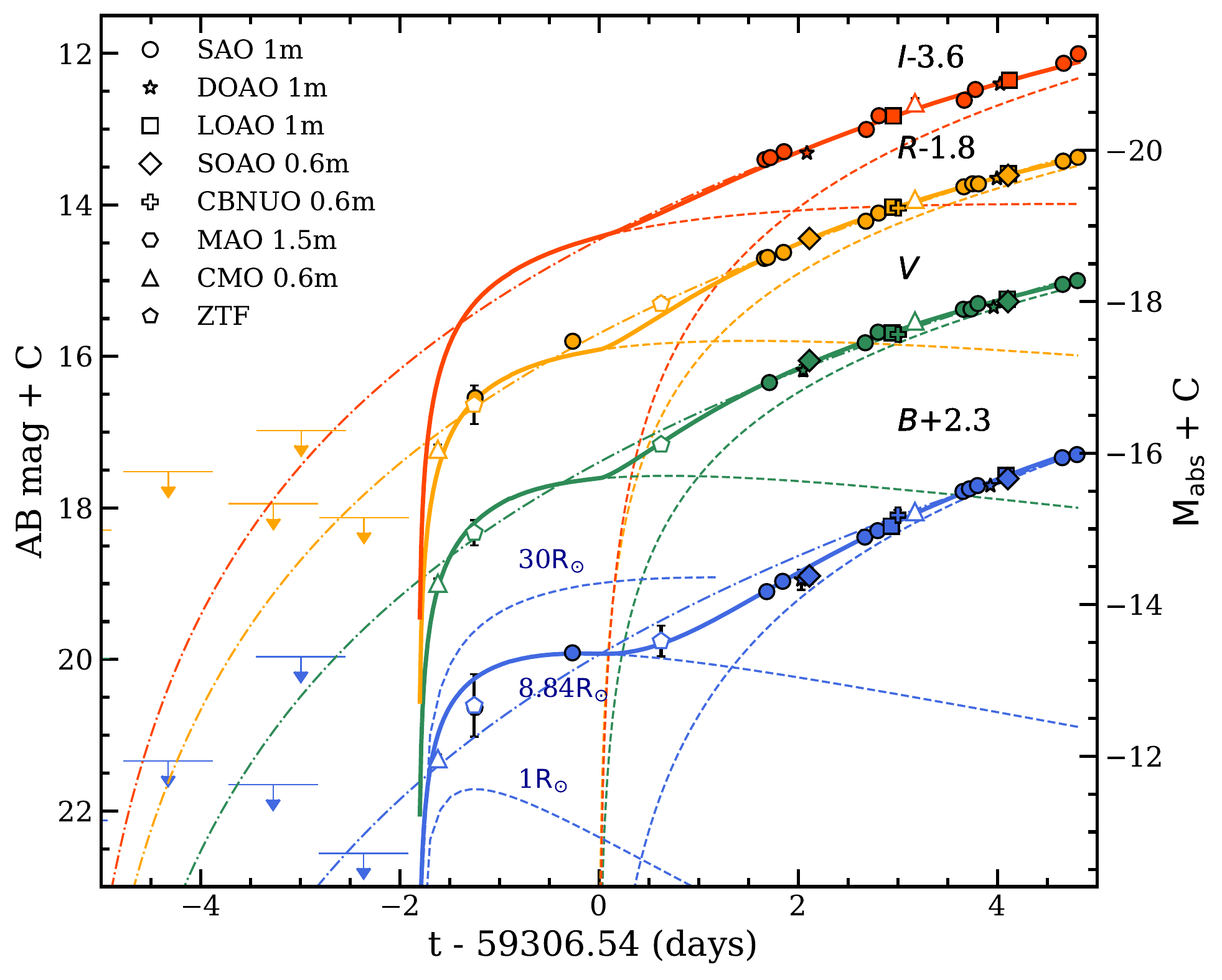}
        \centering
                \includegraphics[width=0.8\textwidth]{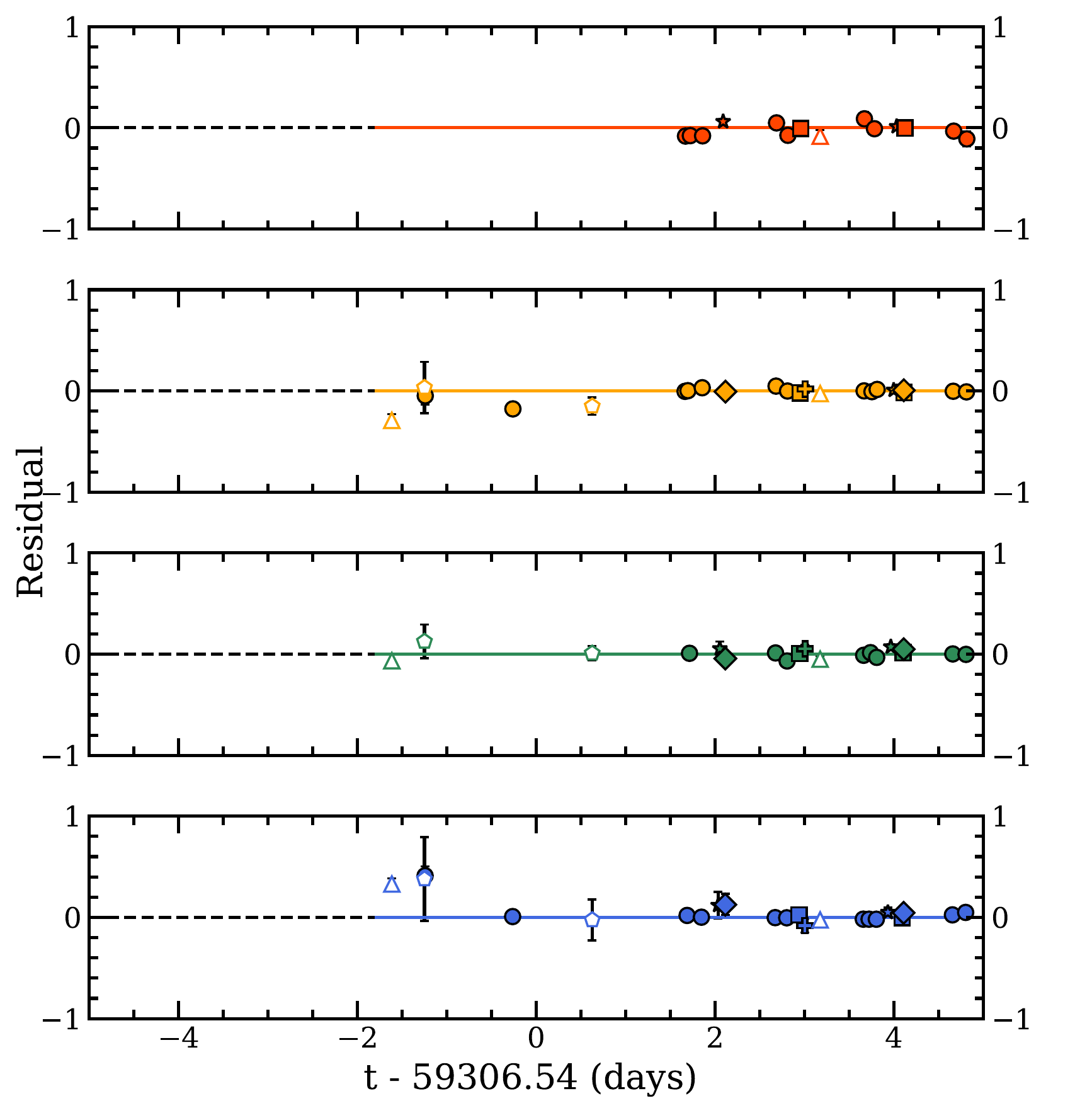}
        \end{minipage}
        \caption{(top) The early light curve of SN 2021hpr between -5 and 5 days from the first light time with the best fit of the two-component model (Solid line). Dashed lines show separated model lines of K10 and the power-law fitting results. K10 models for $1\,R_{\odot}$ and $30\,R_{\odot}$-sized companion stars are also overplotted (Blue dashed lines). Pure simple power-law model lines are also overplotted in dash-dotted lines. Each symbol is the same as that in Figure 3. (bottom) A residual plot corresponding to the top.}  
        \end{figure*}

        \subsection{Early color evolution}
            Our multi-color, high cadence monitoring observation allows us to construct the color curve from the time shortly after the explosion. The 2nd and 3rd left panels of Figure 6 shows the color curve in $B-V$ and $B-R$, for several $R_{*}$ values for the power-law + K10 model. SN 2021hpr was blue very early on, then reddened, and became blue again a few days after the first light time. This overall behavior is in qualitative agreement with our two-component model including SHCE. According to the two-component model, $T_{\rm eff}$ of SHCE increases with $R^{1/4}$, meaning that the larger the companion is, the bluer the early color curve is. Therefore, in this model, if the companion star is large ($R_{*}\sim30\,R_{\odot}$), the predicted colors are very blue. On the other hand, the peak of the color curve in early epochs becomes too red if the companion star is too small. Figure 6 shows the early peak colors agree with a rather small companion star case ($R_{*}\sim9\,R_{\odot}$).

            Additionally, Figure 6 shows the cases where the power-law model is replaced with a delayed detonation models \citep[DDC;][]{1991A&A...245..114K, 2013MNRAS.429.2127B} and pulsational detonation models \citep[PDD;][]{2014MNRAS.441..532D}. These specific models do not include the emission from the ejecta interaction with a companion star, and can possibly fit the early light curve if the SHCE component is included. The SHCE component is taken as the one that best fits the observed data.

            DDC models of \citet{2013MNRAS.429.2127B} are controlled mainly by the transition density $\rho_{\rm crit}$ at which the deflagration is artificially turned into a detonation. Their DDC10 model mimics the light curve of SN 2021hpr after the very early phase, it is chosen for Figure 6. For a similar reason, we plot the PDDEL4n model of \citet{2014MNRAS.441..532D} in Figure 6, which is also know to reproduce SN 2011fe properties.

            When combined with the SHCE model, both DDC10 and PDDEL4n models reproduce the early light and color curves behavior qualitatively, although the PDDEL4n model seems quantitatively deviate from the observed data. We stress that the comparison is done without any sophisticated fitting procedure, so the PDDEL4n model may provide a reasonable fit to the data when some of the SHCE parameters are adjusted.

            We also present other possible results with different configurations in Figure A.1 including the cases of the Power-law+K10 fit at a common viewing angle and other variants of DDC models from \citet{2013MNRAS.429.2127B} and \citet{2014MNRAS.441..532D}. The conclusion we draw from Figure A.1. is similar to Figure 6 that these other models can qualitatively mimic the early light curve behavior with the SHCE component.

        \begin{figure*}\label{fig:fig6}
            \centering\offinterlineskip
            \includegraphics[width=1.1\linewidth]{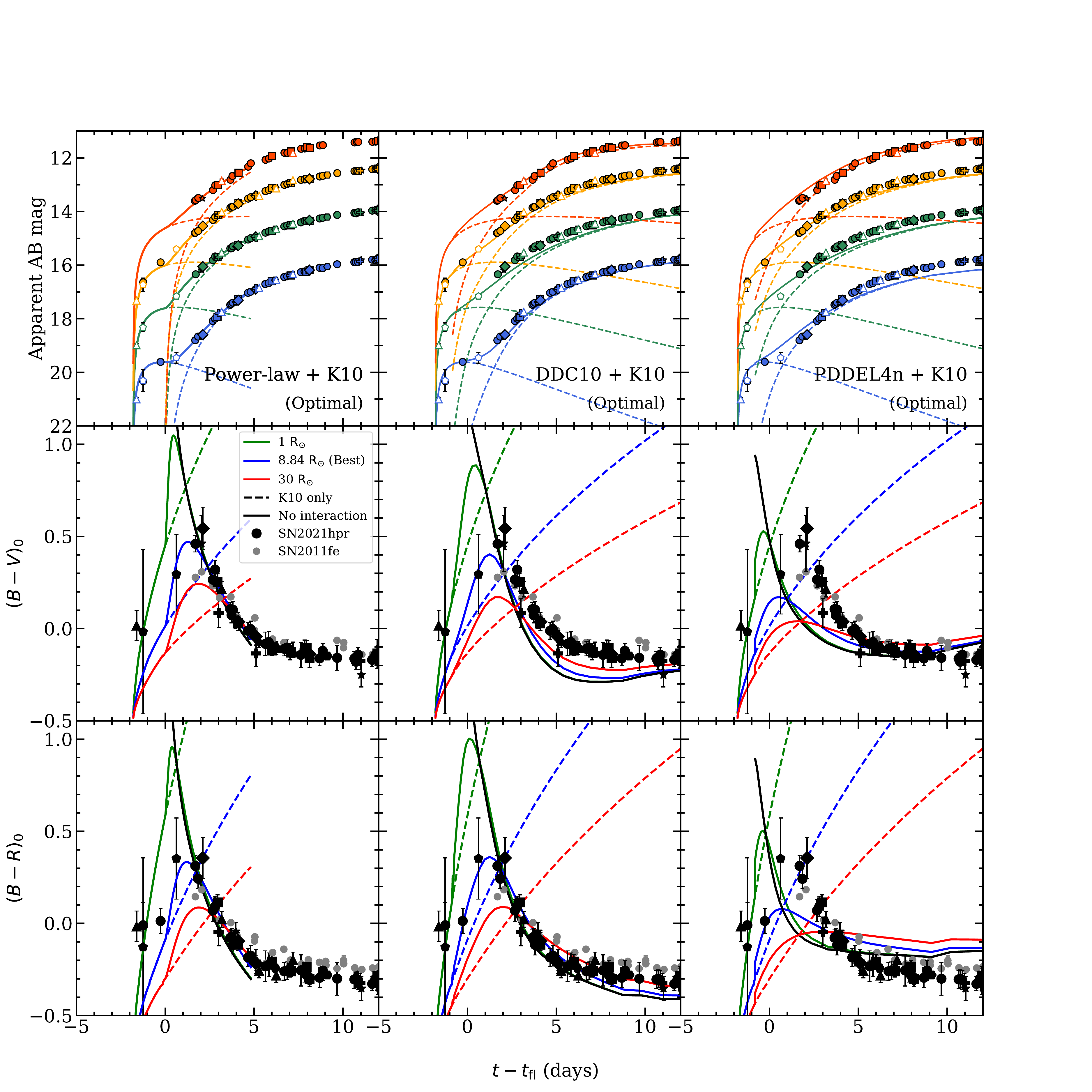}
            \caption{The reddening-corrected light curves (Top row), $(B-V)_{0}$ color (Middle row), and $(B-R)_{0}$ color (Bottom row) curves of SN 2021hpr and SN 2011fe (Black and grey filled circles) in the early phase with the different two-component models (The best fit of Companion+Simple power-law, DDC10 \citep{2013MNRAS.429.2127B}, and PDDEL4n \citep{2014MNRAS.441..532D} at the optimal viewing angle. Two-component models with different companion radii and only K10 models are also presented in solid and dashed lines.}
        \end{figure*}

    \subsection{Optical spectra and the nature of SN 2021hpr}
        To confirm the SN 2021hpr classification, we examined the time evolution of the spectra. Figure 7 shows the optical spectra of SN 2021hpr. For better classification, we additionally include high-quality spectra from the $2.16$-m telescope at XingLong Observatory (XLT) and Transient Name Server\footnote{\url{https://sandbox.wis-tns.org/object/2021hpr}} (TNS) published in \citet{2022PASP..134g4201Z}. We also overplotted the spectrum of SN 2011fe \citep{2013A&A...554A..27P} to compare these SNe. The SAO, XLT, and the 2nd TNS spectra ($-13.9\rm\,$days before the peak brightness) were binned to 10, 3, and 3 pixels respectively to increase their signal-to-noise ratios (SNRs) after 3 sigma clipping. No binning was applied to the other spectra. For the XLT data, we excluded noisy regions at $<3900\,\mathrm{\AA}$.

        \citet{2022PASP..134g4201Z} demonstrated that the spectra of SN 2021hpr show many broad features originated from the intermediate mass elements (IMEs) such as \ion{Si}{2}, \ion{Mg}{2}, High-velocity \ion{Ca}{2} absorption lines. Fe features are also seen but unburned carbon feature (\ion{C}{2}) is not prominent, which is seen in SN 2011fe. The very high expansion velocity of \ion{Si}{2} decreases at rates of about $-800\,\mathrm{km/s}$ per day making SN 2021hpr classified as the high velocity gradient group (HVG) SNe Ia. In the SAO spectra, we can also find the IME features ranging from $4600\,\mathrm{\AA}$ to $9000\,\mathrm{\AA}$ despite their poor SNRs. Applying GEneric cLAssification TOol \citep[GELATO;][]{2008A&A...488..383H} on the April 14 spectrum (taken at near maximum brightness), we find that the SN 2021hpr spectrum is similar to SN 1989B, a normal SN Ia. Its width-luminosity relation (Figure 4), the similarity of its light curve to SN 2011fe, and the spectral features all suggest that SN 2021hpr is a normal SN Ia.

        We discuss SNe Ia with or without early flux excess in terms of spectral diversity with some cases. Early flux excess is found in some luminous SNe Ia (99aa-like) showing weak or no \ion{Ca}{2} and \ion{Si}{2} absorption features such as SN 2015bq \citep{2022ApJ...924...35L}, and iPTF14bdn \citep{2015ApJ...813...30S}. Likewise, a sub-luminous SN Ia, iPTF14atg \citep{2015Natur.521..328C} is reported to have early flux excess in the ultraviolet wavelength. MUSSES1604D \citep{2017Natur.550...80J} is a hybrid SN Ia classified as a normal SN Ia in photometry but has a peculiar \ion{Ti}{2} absorption feature in the spectrum. SN 2017cbv is close to a normal SN Ia with transitional characteristics such as weak \ion{Si}{2} and \ion{Ca}{2} absorption features but stronger than those of 99aa-like SNe Ia \citep{2017ApJ...845L..11H}. On the other hand, normal SNe Ia with early flux excess are also discovered such as SN 2012cg \citep{2016ApJ...820...92M}, SN 2018aoz \citep{2022NatAs.tmp...45N}, SN 2018oh \citep{2019ApJ...870...12L}, and SN 2021hpr in this paper.

        Unburnt carbon features (e.g. \ion{C}{2} $\lambda6580$) can be found in both SNe Ia with (SN 2012cg, SN 2017cbv, SN 2018oh and iPTF14atg) and without early flux excess (SN 2011fe; \citealt{2013A&A...554A..27P}, SN 2012ht; \citealt{2014ApJ...782L..35Y}, and SN 2013dy; \citealt{2013ApJ...778L..15Z}). High velocity features (HVFs) near the maximum brightness seem to appear in both kinds of SNe Ia with (SN 2021hpr; \citealt{2022PASP..134g4201Z}, SN 2021aefx; \citealt{2022ApJ...933L..45H}) and without early flux excess (SN 2012fr; \citealt{2014AJ....148....1Z, 2018ApJ...859...24C}, SN 2019ein; \citealt{2020ApJ...893..143K}). Further spectroscopic data, especially obtained at the early time, is required to understand the relation between spectral features and the early flux excess.

        \begin{figure*}\label{fig:fig2}
            \centering
            \includegraphics[width=1.\textwidth]{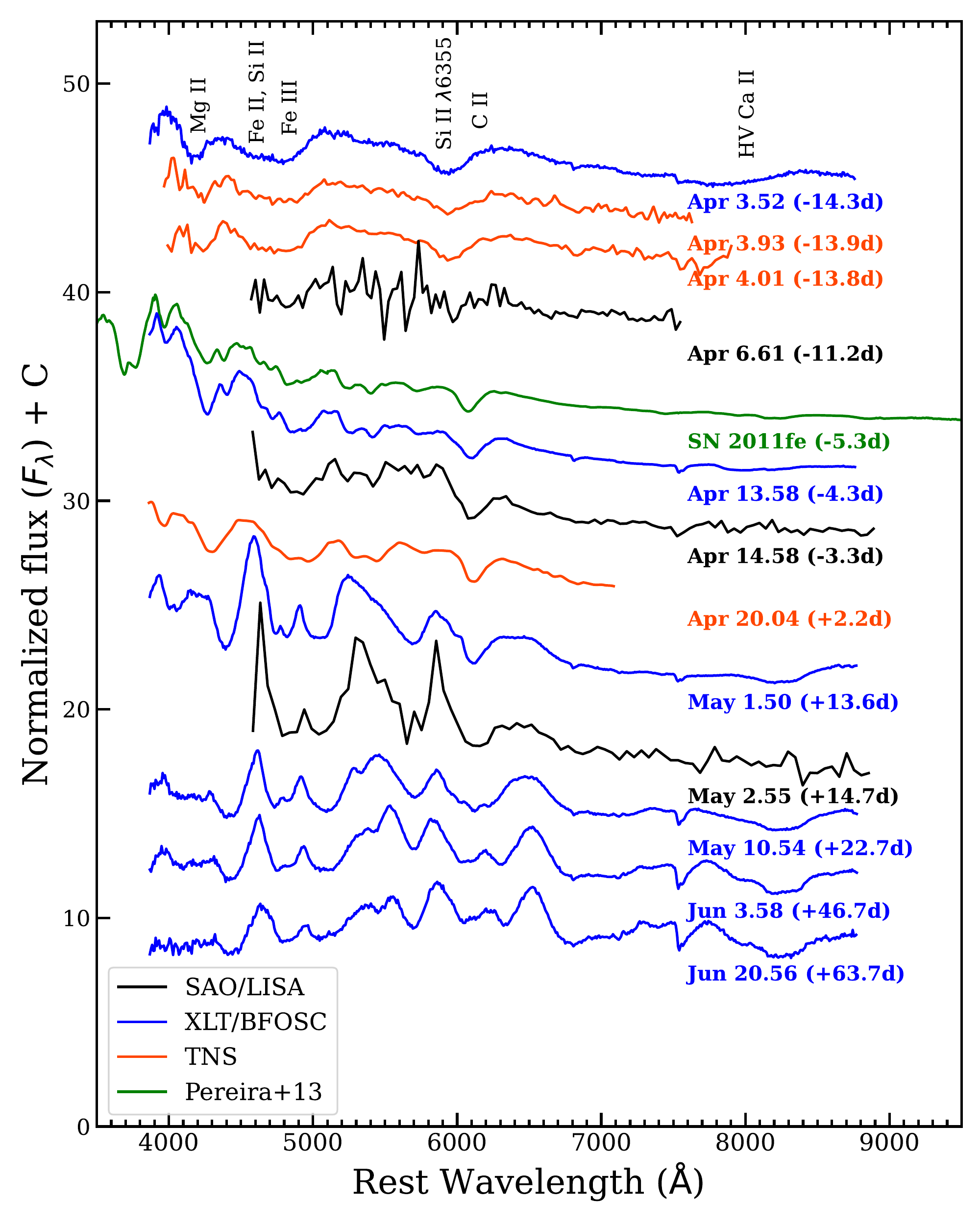}
            \caption{Optical spectral evolution of SN 2021hpr observed from the multiple instruments including SAO (Black), XLT (Blue), TNS (Orange-red). Normalized fluxes are shifted with arbitrary constant values. No extinction is corrected except SN 2011fe (Green).}
        \end{figure*}

        \subsection{Finding Possible Progenitor System in the Pre-explosion HST image}
            We can possibly constrain the progenitor system, especially for the companion star, by directly identifying it at the SN position in the pre-explosion images \citep{2011Natur.480..348L, 2014Natur.512...54M}. We identified a series of Hubble Space Telescope (HST) images from the HST archive\footnote{\url{https://archive.stsci.edu/}} taken before the SN explosion during November 2017 to March 2018 (Proposal 15145; PI: A. Riess) and after the explosion (Proposal 16691, PI: R. Foley). The images were obtained by Wide-Field Camera 3 (WFC3) in F350LP, F555W, F814W, and F160W filters. Table 6 summarizes the observation. The single frame images were stacked using \texttt{Swarp} \citep{2010ascl.soft10068B}. Figure 8 shows the HST images before and after the SN explosion. The coordinate of SN 2021hpr and its 1-$\sigma$ error, $0\farcs3$ from Gaia alerts \citep{2019TNSAN..60....1Y} in TNS is drawn as circle in the figure.

            At the SN 2021hpr position, no obvious source was found in the pre-explosion image. We measured a $3\sigma$ detection limit for a point source with the default aperture size of $0\farcs2$ radius, finding upper limits on the progenitor system magnitudes of $\sim27-28\,$mag in optical, and $\sim25\,$mag in F160W (Table 6).

            \begin{figure*}
                \includegraphics[width=1.\textwidth]{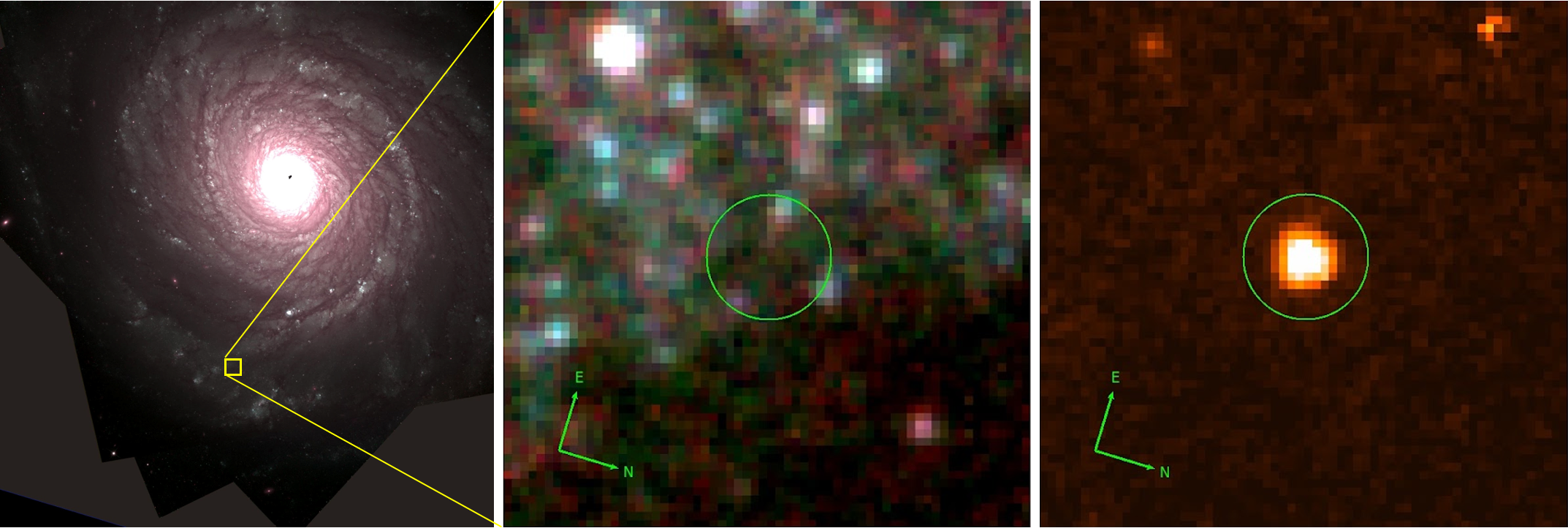}
                \centering
                \caption{(Left) An F350LP/F555W/F814W color image before the explosion of SN 2021hpr observed by the Hubble Space Telescope/Wide Field Camera 3 (WFC3). (Middle) A zoomed region ($2\farcs5\times2\farcs5$) of the region in the yellow box of the full-frame image on the left. The site of SN 2021hpr is marked with a green circle centered at the \textit{Gaia} alert coordinated with a radius of the $1\sigma$ astrometric accuracy ($0\farcs3$). (Right) The detection image of SN 2021hpr in the F814W filter on Dec 29th, 2021. FOV is the same with the middle panel. The source on the upper right is an artifact.} 
            \end{figure*}

            Figure 9 shows the color-magnitude diagram (CMD) with stellar evolutionary tracks and HST upper limits. The tracks are calculated from the MESA Isochrones and Stellar Tracks \citep[MIST;][]{2016ApJ...823..102C}, a recent set of stellar evolutionary tracks and isochrones, which provides the synthetic photometry in HST/WFC3 filters\footnote{\url{http://waps.cfa.harvard.edu/MIST/index.html}}. We adopted the tracks of initial mass ($M_{\rm init}$) from $8-16\,M_{\odot}$ with a step of $2\,M_{\odot}$ and solar metallicity. We also plotted the Bessell $V$- and $I$-band synthetic photometry\footnote{The synthetic photometry values were calculated from the spectral energy distribution (SED) fit of samples in the literature. We plotted the Bessell filter magnitudes because there is no large difference between the $V$-, $I$-band, and F555W-, F814-band magnitudes.} of some evolved stars including asymptotic giant and supergiant branch in the Large Magellanic Cloud (LMC) \citep[black filled circles;][]{2018A&A...609A.114G}.

            In the CMD, stars with $M_{\rm init}\gtrsim10\,M_{\odot}$, except for some in high luminosity phase, would have been detected in the HST image. The radius of $M_{\rm init} \sim 10\,M_{\odot}$ star can be approximated with the evolutionary tracks in \citet{2005ApJ...628..973L}. Assuming $\rm log_{10}(T_{\rm eff}[\rm K])=3.57$, $M_{\rm bol}=-5.0 \rm\,mag$ as the effective temperature and bolometric magnitude of stars with $M_{\rm init}=10\,M_{\odot}$, we obtain ${L/L_{\odot}}\sim7870$ giving us $R_{*}\sim215\,R_{\odot}$ as an upper limit of the radius of the companion star.

        \begin{deluxetable}{l|cccc|c}\label{tbl:table6}
            \tablecaption{Description of HST WFC3/UVIS and IR imaging images. Host reddening ($A_{\lambda, \rm host}$) is calculated at the pivot wavelength. $M_{\rm abs,0}$ is the Milky Way and host reddening corrected absolute magnitude.}
            \tablecolumns{6}
            \tablenum{6}
            \tablewidth{0pt}
            \tablehead{
            \colhead{} &
            \multicolumn{4}{|c|}{Pre-SN}&
            \multicolumn{1}{r}{Post-SN}
            }
            \startdata 
                Filter & F350LP & F555W & F814W & F160W & F814W \\
                Detector & UVIS & UVIS & UVIS & IR & UVIS\\
                $t_{\rm exp}$ (s) & $25520$ & $5952$ & $5954$ & $12055$ & $780$ \\
                $N$ of images & $11$ & $5$ & $5$ & $5$ & $1$\\
                Pivot $\lambda$ ($\rm \AA$) & $5862.5$ & $5308.2$ & $8034.2$ & $15369.2$ & $8034.2$\\
                $3\sigma$ limit (AB) & $>28.10$ & $>28.04$ & $>27.45$ & $>25.41$ & $>26.56$ \\
                $A_{\lambda, \rm host}$ (mag) & $0.27$ & $0.31$ & $0.16$ & $0.06$ & $0.16$\\
                $M_{\rm abs, obs}$ (AB) & $>-5.18$ & $>-5.24$ & $>-5.83$ & $>-7.87$ & $>-6.72$ \\
                $M_{\rm abs,0}$ (AB) & $>-5.45$ & $>-5.56$ & $>-6.00$ & $>-7.92$ & $>-6.89$
            \enddata 
        \end{deluxetable}

        \begin{figure}\label{fig:fig9}
            \centering
            \includegraphics[width=1\linewidth]{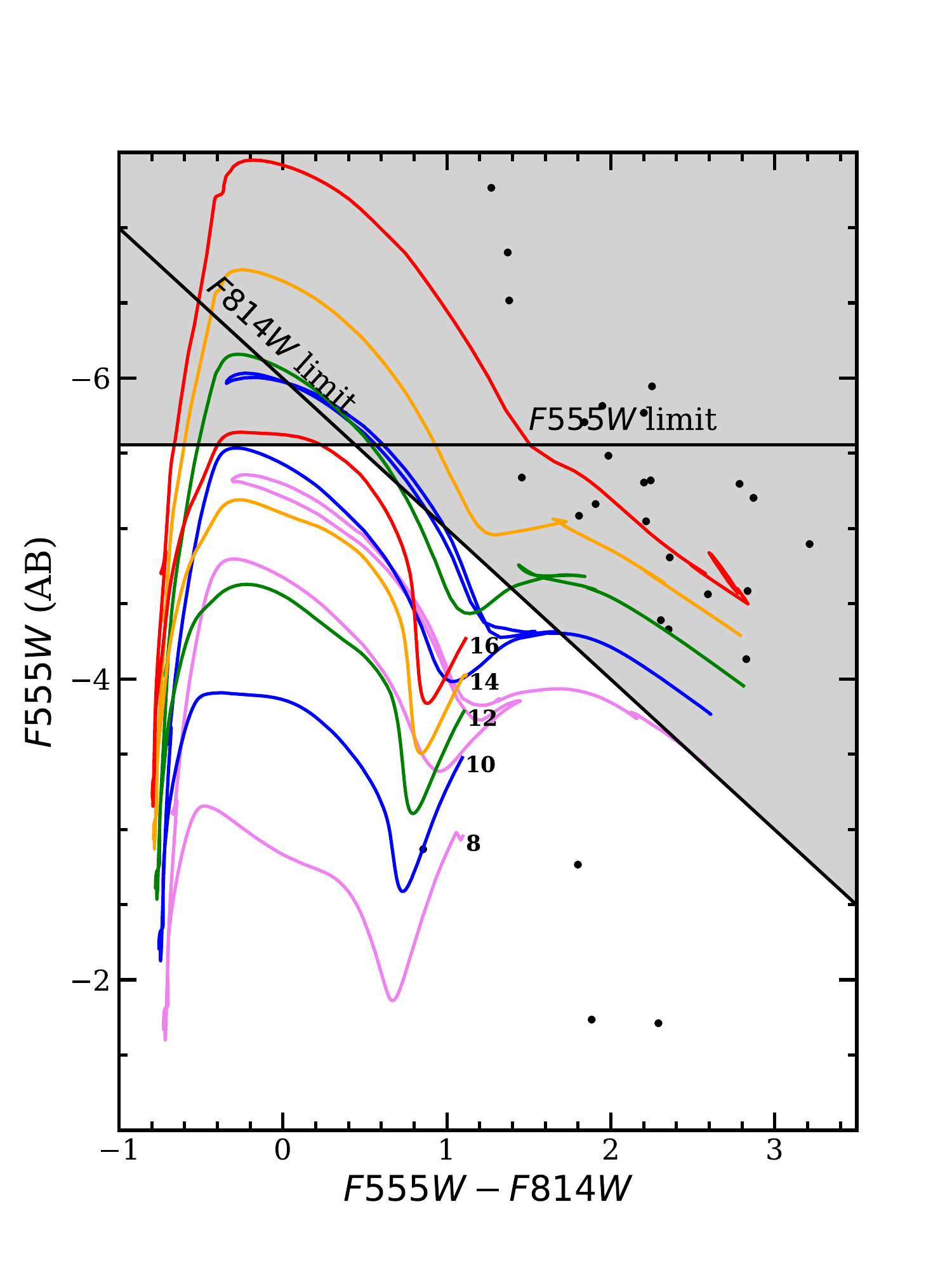}
            \caption{A color-magnitude diagram of HST filters with MIST evolutionary tracks for massive stars with initial masses from $8-16\,M_{\rm \odot}$ assuming the solar metallicity. The initial masses are marked at the starting points of the tracks. The gray shaded area shows an excluded parametric space for the progenitor system based on the \textit{HST} non-detection. Extinction correction is applied to derive the upper limits here (the black solid lines). The Bessell synthetic $V$- and $I$-band photometry of evolved stars in LMC are marked as black-filled circles.}
        \end{figure}

        \subsection{The stripped mass limit from HET late phase spectroscopy}
            To identify emission lines from the stripped matter of the companion, we also obtained an optical spectrum of SN 2021hpr using the blue pair of the second generation Low-Resolution Spectrograph (LRS2-B) mounted on the $9.2\,\rm m$ Hobby-Eberly Telescope at McDonald Observatory, USA \citep{2014SPIE.9147E..0AC}. LSR2-B is a $12''\times7''$ Integral Field Unit (IFU) that covers the wavelength ranges of $3700 \leq \lambda\,(\AA) \leq 4700$ (R$\sim1900$) and $4600 \leq \lambda\,(\AA) \leq 7000$ (R$\sim1100$). A single frame of 1000\,seconds was obtained under a dark condition ($g\sim21.13$\,mag arcsec$^{-2}$) on 2021 November 30.26 UT. At that time, SN 2021hpr was in a late phase ($59548.26 \rm \,MJD$, +226\,days from the $B$-maximum brightness).

            The spectrum was reduced using the code Panacea\footnote{\tt https://github.com/grzeimann/Panacea}, the standard pipeline of HET LRS2. Flux calibration was conducted by observing HD 55677 as a spectrophotometric standard and using this spectrum to set the zero point of the response curve, while the shape of the response curve was constructed from a sequence of standard stars observed over 6 months in 2019. We used the redshift of $z=0.009346$ from \citet{2021TNSCR1031....1T} to shift the spectrum to the rest-frame. Milky Way and host galaxy extinctions were also corrected. Furthermore, we re-calibrated the flux of the spectrum so that the flux values of the spectrum match our photometry at the observed date by multiplying $1.9$ on the flux as a correction factor. It is not clear why the integrated flux from the HET spectrum is different from the value from the image photometry. Varying weather condition could be the reason.

            Figure 10 shows the reduced spectrum. To search for nebular emission lines, we subtracted the supernova nebular flux features in the following way. We adopted the method from \citet{2019ApJ...872L..22T} for the nebular flux fit. We first masked regions around spectral lines such as the Balmer series lines (H$\alpha$, H$\beta$, H$\gamma$), \ion{He}{1} $\lambda\,5876$, and \ion{He}{1} $\lambda\,6678$ with the width of 1000\,km/s ($W_{\rm line}$) that is known to be the line width of the stripped matter \citep{2000ApJS..128..615M,2017MNRAS.465.2060B}. Then, the spectrum was smoothed using the 2nd order Savitzky-Golay polynomial \citep{1992nrfa.book.....P} with a window size of 3000\,km/s in $6563\,\AA$ which is wider (narrow) than the host galaxy (the ejecta) features. Considering $R=1100$ at \ion{He}{1} $\lambda\,6678$, the observed data were binned to a wavelength size of 6\,$\rm\AA$. The observed data and the fitted nebular flux are shown as the black and red lines in Figure 10.

            \begin{figure*}[t]\label{fig:fig10}
                \centering
                \includegraphics[width=1\textwidth]{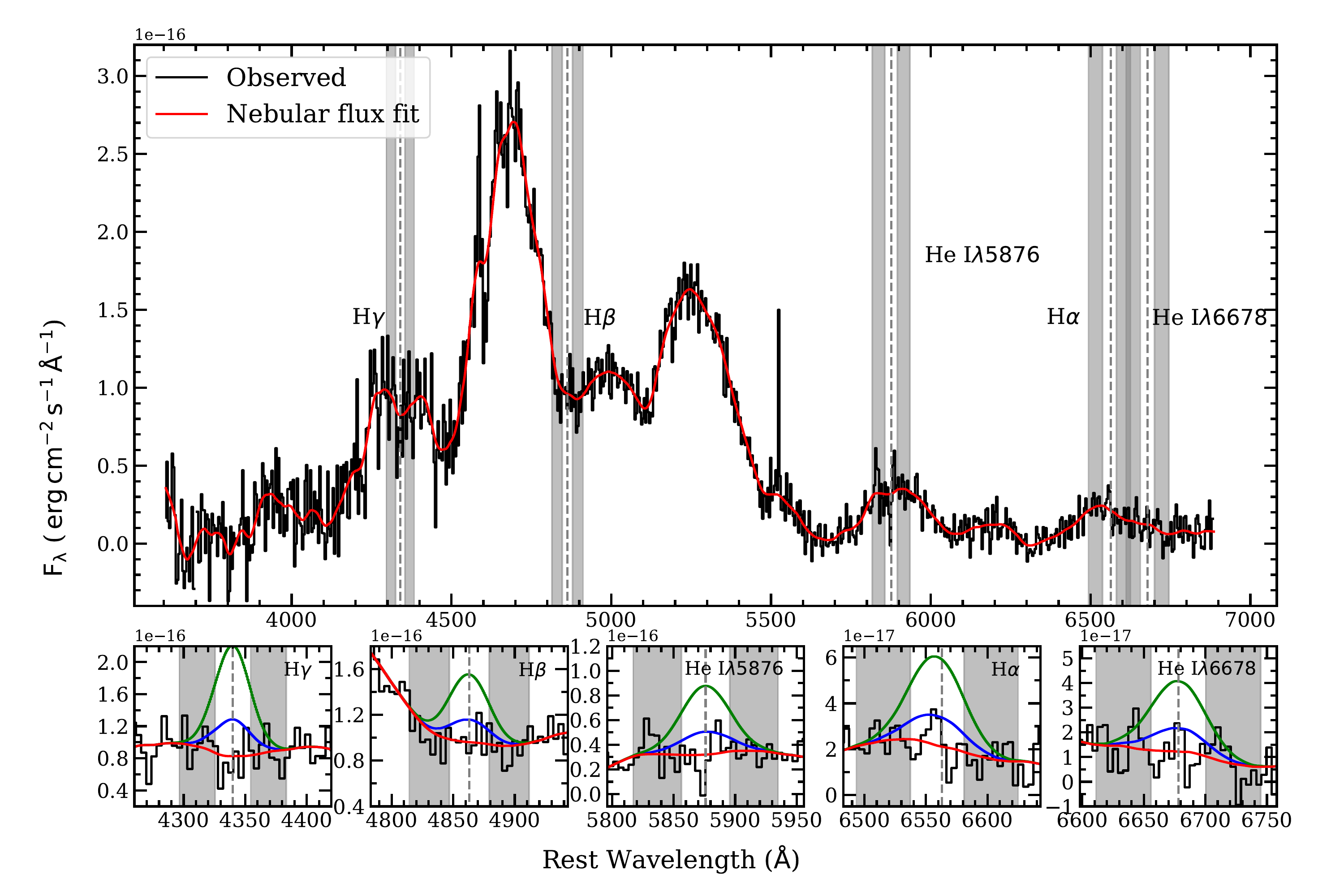}
                \caption{Flux-calibrated HET/LRS2-B spectrum of SN 2021hpr at the nebular phase of $+243\,$days since the explosion (black solid line) and the nebular flux fit (red solid line). The gray dashed lines mark the positions of each spectral line. Gray shaded areas are the regions for measuring the RMS of the nebular flux. The bottom five panels show spectral regions around nebular emission lines and the $1\sigma$- and $3\sigma$-line flux limits (blue and green solid lines, respectively). Fluxes are binned at a size of $6\,\rm \AA$.}
            \end{figure*}

            \begin{deluxetable*}{cccccc}
            \tablecaption{The RMS of the nebular features and flux and luminosity limits of nebular emission lines, along with the upper limits on the stripped mass corrected by multiplying the correction factors expected in +250\,days since explosion.}
            \tablenum{7}
            \tablehead{
            \colhead{Line} &
            \colhead{$W_{\rm {line}}$} &
            \colhead{Nebular flux noise ($\sigma$)} &
            \colhead{Flux limit ($3\sigma$)} &
            \colhead{Luminosity limit ($3\sigma$)} &
            \colhead{$M_{\rm {st}}$ limit} \\
            \colhead{} &
            \colhead{($\rm {\AA}$)} &
            \colhead{($\rm {10^{-17}\,erg\,\AA^{-1}\,cm^{-2}\,s^{-1}}$)} &
            \colhead{($\rm {10^{-16}\,erg\,\AA^{-1}\,cm^{-2}\,s^{-1}}$)} &
            \colhead{($\rm {10^{38}\,erg\,s^{-1}}$)} &
            \colhead{($M_{\odot}$)}
            }
            \startdata 
                H$\alpha$                    & $21.89$ & $1.29$ & $8.48$ & $2.08$ & $<0.003$ \\
                H$\beta$                     & $16.22$ & $1.97$ & $9.60$ & $2.35$ & $<0.003$ \\
                H$\gamma$                    & $14.48$ & $4.60$ & $19.99$ & $4.89$ & $<0.005$ \\
                \ion{He}{1}{$\lambda\,5876$} & $19.60$ & $1.86$ & $10.93$ & $2.68$ & $<0.003$ \\
                \ion{He}{1}{$\lambda\,6678$} & $22.28$ & $0.95$ & $6.37$ & $1.56$ & $<0.002$ \\
            \enddata
            \end{deluxetable*}

            After subtracting the best-fit nebular flux from the observed spectrum, we searched for signs of emission lines in the corresponding spectral regions, but no significant emission lines were found (Figure 10). We measured emission line flux limits from the nebular flux RMS around each line. The RMS was measured as the standard deviation of the Gaussian fit of the nebular flux distribution ranging from $-3\times W_{\rm line}$:-$W_{\rm line}$ to $W_{\rm line}$:$3\times W_{\rm line}$, excluding the signals at the position of each emission line (The gray shaded regions in Figure 10). The $1 \sigma$ and $3 \sigma$ flux limits are also plotted together in Figure 10 assuming a Gaussian profile with a width of $1000\,\rm km/s$. These flux limits are converted to luminosity limits considering the distance. For the H$\alpha$ emission line, the $3\sigma$ luminosity limit is $2.08\times10^{38}\,\rm erg/s$. Using the H$\alpha$ luminosity-stripped mass relation of the MS38 model (Equation 1 from \citealt{2018ApJ...852L...6B}), we estimate the $3\sigma$ stripped mass limit ($M_{\rm st}$) from each emission line\footnote{We adopted the form provided in \citet{2018ApJ...863...24S} as Equation (1) in \citet{2018ApJ...852L...6B}.}. Since this model gives a prediction at $200$\,days after the explosion, we estimated a scale factor to calculate $M_{\rm st}$ at $+243$\,days since the explosion. This was done by adopting the luminosity ratio between 200 and 243 days as the scaling factor, since the ratio of bolometric luminosity to H$\alpha$ luminosity is known to be constant \citealt{2018ApJ...852L...6B}. During $200$ and $250$\,days since the explosion, SN 2021hpr was observed in $BR$ bands. In this time range, $B$- and $R$-band flux have decreased by a factor of 0.57 and 0.77 with little $B-R$ color change suggesting that $B$ and $R$ bands luminosities change roughly like the bolometric luminosity. Hence, we multiplied 0.67 (the mean of 0.57 and 0.77) to the model value at 200 days to convert it to the value at 243 days. After the correction, the $3\sigma$ stripped mass limit ($M_{\rm st}$) for H$\alpha$ is $<0.003\,M_{\odot}$. For the other Balmer lines, the mass limits are also presented in Table 6. For He lines, their mass limits are obtained assuming that the luminosities of He lines follow Equation (1) of \citet{2018ApJ...852L...6B}.

            For a He star companion, hydrogen lines would not be visible. Yet, a stripped mass of $\lesssim 0.06\,M_{\odot}$ is expected, and the predicted strengths of the He lines are only a factor of a few smaller than the hydrogen lines in H-rich companion star model \citep{2018ApJ...852L...6B}. No strong He emission lines in our data suggest a small amount of stripped He mass.

    \section{Discussion} \label{sec:Discussion}
        \subsection{SD System as SN 2021hpr Progenitor}
            In Sections 3.3 and 3.4, we have shown that the companion interaction model can explain the early light excess and its color evolution of SN 2021hpr. The result suggests the possibility of an SD system with a $\sim9\,R_{\odot}$ companion star as the progenitor system of SN 2021hpr. A $\sim9\,R_{\odot}$ companion can be a subgiant star with $6\,M_{\odot}$ or low mass red giant \citep{1996ApJ...470L..97H}. On the other hand, the radius of $\sim9\,R_{\odot}$ is too large for a low mass main-sequence star. A He-rich envelope star (He star) can also be a companion because its orbital separation $a$, assuming a circular orbit, ranges from $4-80\,R_{\odot}$ \citep{1999ApJ...519..314H} with $a=2-3\,R_{*}$ for typical mass ratios \citep{1996ApJ...470L..97H, 2010ApJ...708.1025K}.

            However, this interpretation needs to be reconciled with no signatures of H$\alpha$ emission in the late spectrum, since we expect to see strong nebular emission lines in a late phase for the SD model. We provide several possible ways to explain the no detection of the nebular lines.

            Several works note that the stripped mass is reduced if the binary separation distance is large \citep{2000ApJS..128..615M, 2008A&A...489..943P, 2012A&A...548A...2L, 2012ApJ...750..151P, 2017MNRAS.465.2060B}. \citet{2008A&A...489..943P} demonstrates this in their Equation (4). \citet{2008A&A...489..943P} show $M_{\rm st} \sim a^{-3.5}$. Applying this relation to their models, it is not too difficult to obtain the limit of $M_{\rm st} < 0.01\,M_{\odot}$. For example their rp3\_24a model, where the companion star's initial mass is $2.4\,M_{\odot}$ and the separation of the binary system is $4.39 \times 10^{11}\,$cm (or $6.3\,R_{\odot}$), they get $M_{\rm st} \sim 0.01\,M_{\odot}$. Making $a$ a bit further will easily reduce $M_{\rm st}$ to a value less than $0.01\,M_{\odot}$.

            \citet{2008A&A...489..943P} also showed that low explosion energy produces a small amount of the stripped mass (Equation (2) in \citealt{2008A&A...489..943P}), so this could be another reason for the non-detection of H$\alpha$. However, considering SN 2021hpr is a normal SN Ia event, low explosion energy would make SN 2021hpr a sub-luminous event.

            Overall, we conclude from the early multi-band light curve that an SD binary system with a companion star with a stellar radius of $ \sim 9\, R_{\odot}$ can be a progenitor system of SN 2021hpr. However, no or weak nebular emission lines in a late phase pause a challenge to this interpretation. Further investigation on this issue is needed.

        \subsection{Double Detonation Model}     
            As mentioned in the introduction, the early color of SN 2021hpr can be regarded as a ``red bump'' in some DDet models. The DDet model is a model where the thermonuclear explosion in the He shell causes the core ignition. DDet models with a thick He shell are known to produce excess in early light curve due to radioactive materials in the He shell ashes (e.g., \citealt{2019ApJ...873...84P}). On the other hand, the He shell ashes contain a large amount of Fe-group elements that block photons at short wavelengths and make the SNe colors red. Qualitatively speaking, one would expect red excess light in the early light curve in thick He shell DDet models, which is possibly in agreement with SN 2021hpr's color and light curves.

            Figure 11 compares the SN 2021hpr light and color curves with a thick He shell DDet model with $0.9\,M_{\odot}$ WD+$0.08 \,M_{\odot}$ He shell (edge-lit) of \citet{2019ApJ...873...84P}. The shape of the early red peak is similar to the observed colors but the model produces a slower evolution of the red early light curve than observed. Furthermore, the DDet model produces a light curve that is too red at a later time ($15\,$days since explosion in Figure 11). We conclude that DDet models have difficulties reproducing the SN 2021hpr light curve.

        \subsection{Alternative Explosion Scenarios}
            \citet{2020A&A...642A.189M} demonstrate the early flux excess can be produced from the existence of the $^{56}\rm Ni$ clump which depends on its mass, width, and location in the outer ejecta ($^{56}\rm Ni$ clump model). Figure 12 compares the light and color curves of SN 2021hpr with one of the $^{56}\rm Ni$ clump models ($0.005\,M_{\odot}$ $^{56}\rm Ni$ clump with a width of $0.06\,M_{\odot}$ on the fiducial light curve of SN 2018oh). This model is not completely consistent with the light curve. Still, it produces the early red excess peak at $B-V\sim 0.5$ and keeps the color relatively blue at later epochs but perhaps too blue and qualitatively reproduces the observed light curve features. Considering that the model does not require the production of H-Balmer emission lines in the nebular phase, it may provide a possible way to explain the early evolution of SN 2021hpr.

            \citet{2015MNRAS.447.2803L} derived an analytic form of an early signal emitted from the interaction between the SN ejecta and the disk-originated matter around the primary WD. This matter results from the tidal disruption of the companion WD (Disk-originated matter; DOM). This emission is also expected to last up to a few hours in the ultraviolet wavelengths. The light curves of SN 2012cg and iPTF2014atg were also examined by the ejecta-DOM interaction in addition to other suggested models such as stratified $^{56}\rm Ni$ structure, DDet with an outer $^{56}\rm Ni$ shell, and the companion interaction \citep{2017MNRAS.470.2510L}. In their another study, \citet{2019ApJ...872L...7L} argued that the early blue excess of SN 2018oh can be fitted with the two-component DOM interaction model better than the companion model. SN 2018oh is another SN without late-phase H$\alpha$ emission from the stripped matter of its donor star in the SD system \citep{2019ApJ...872L..22T}, so the DOM model may be able to explain the observed properties of SN 2021hpr.

            \citet{2012ApJ...758..123W} pointed out that a strongly magnetized WD and an M dwarf star pairs, which are quite common in the Galaxy, can be the SN Ia progenitor. The material from the M dwarf star can be locked by their combined magnetic fields (``magnetic bottle'') on the magnetic pole of the WD, producing an SN Ia with an extra light source originating from the material from the M dwarf and the accreted matter of the WD. It is not clear if SN 2021hpr can be explained with this model, but it will be interesting to further investigate outcomes from this model.

            \begin{figure*}\label{fig:figure11}
               \begin{minipage}{.32\textwidth}
                 \centering
                 \includegraphics[width=1.\textwidth]{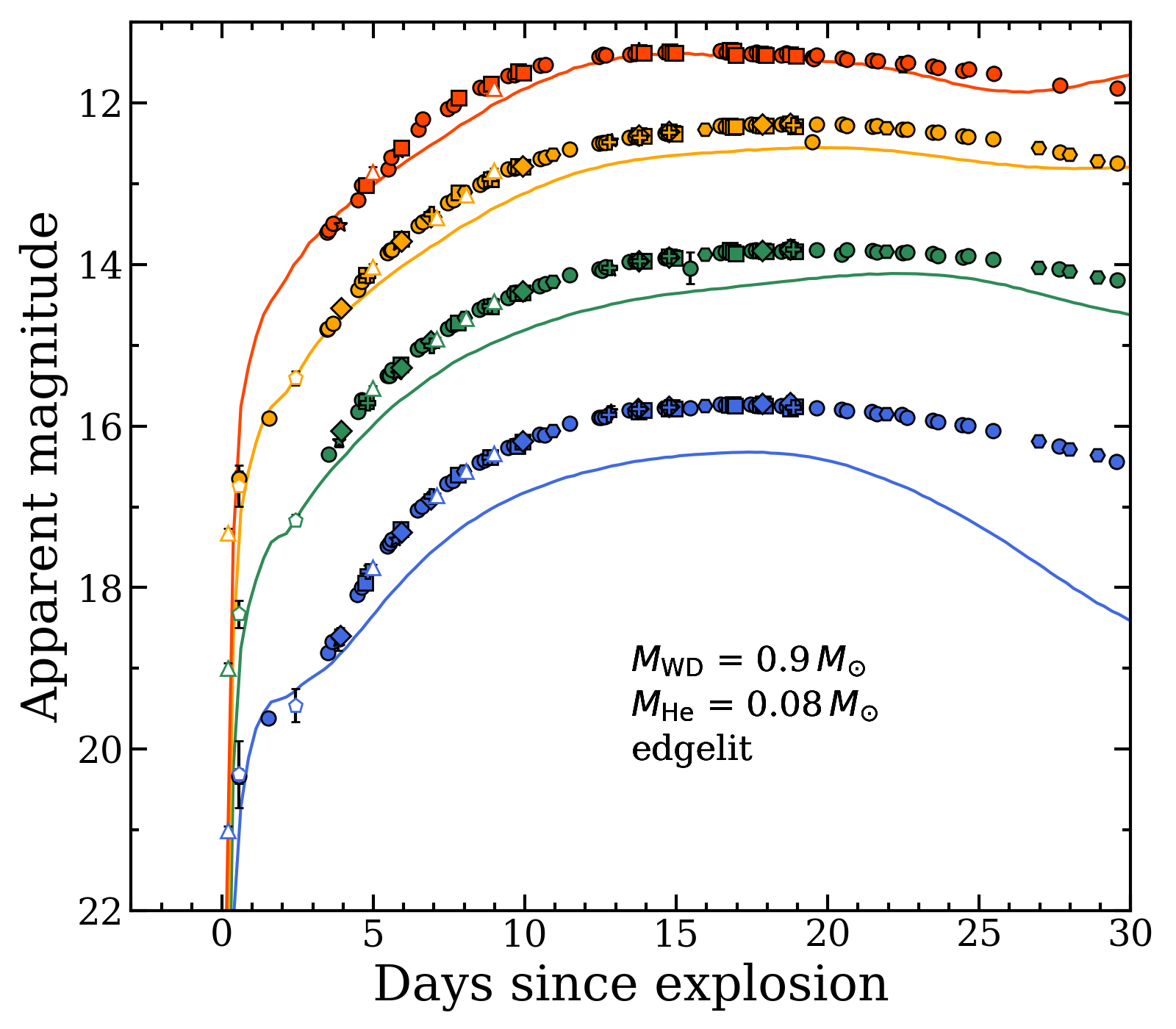}
               \end{minipage}
               \hfill%
               \begin{minipage}{.32\textwidth}
                 \centering
                 \includegraphics[width=1.\textwidth]{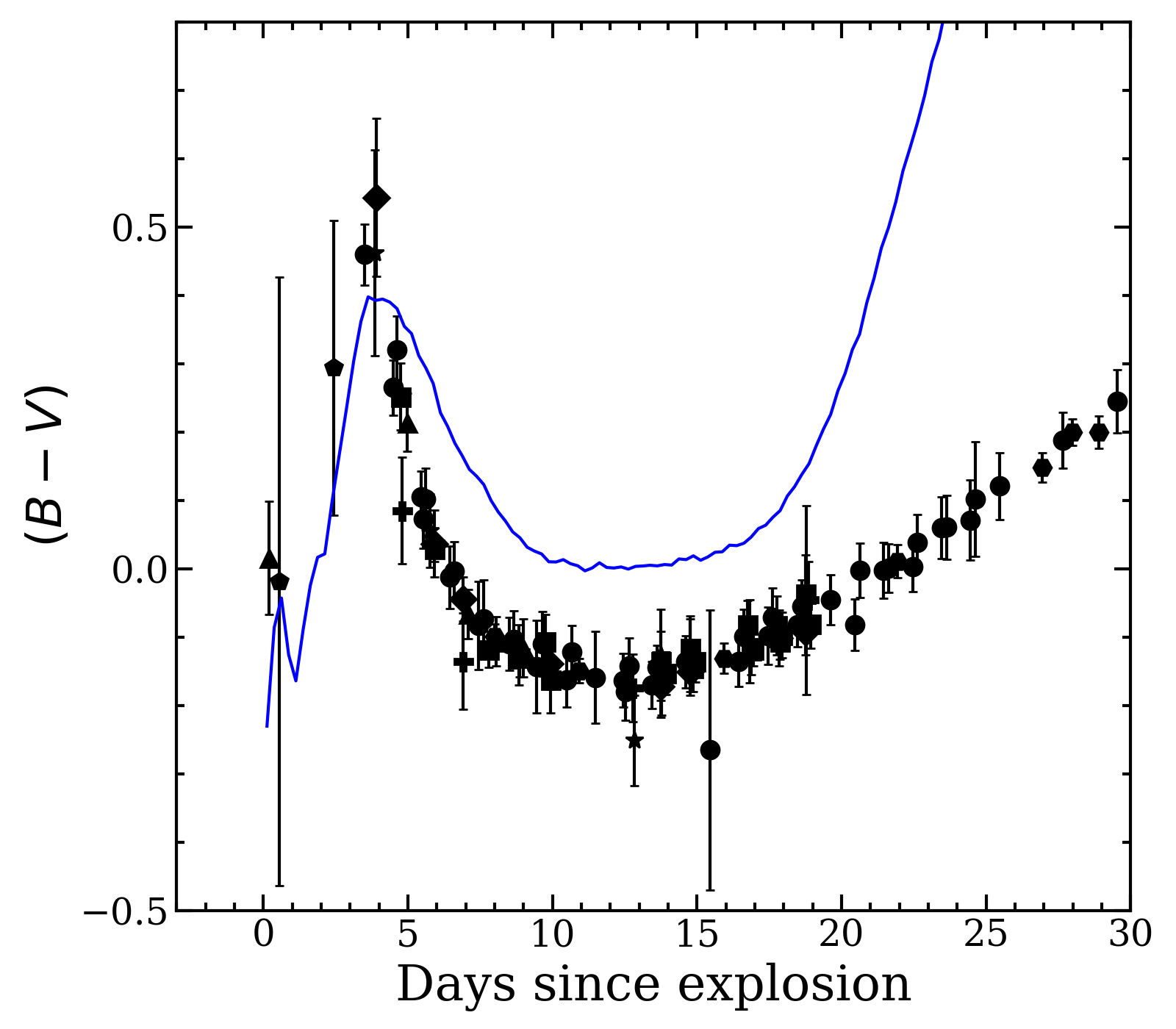}
               \end{minipage}
               \hfill%
               \begin{minipage}{.32\textwidth}
                 \centering
                 \includegraphics[width=1.\textwidth]{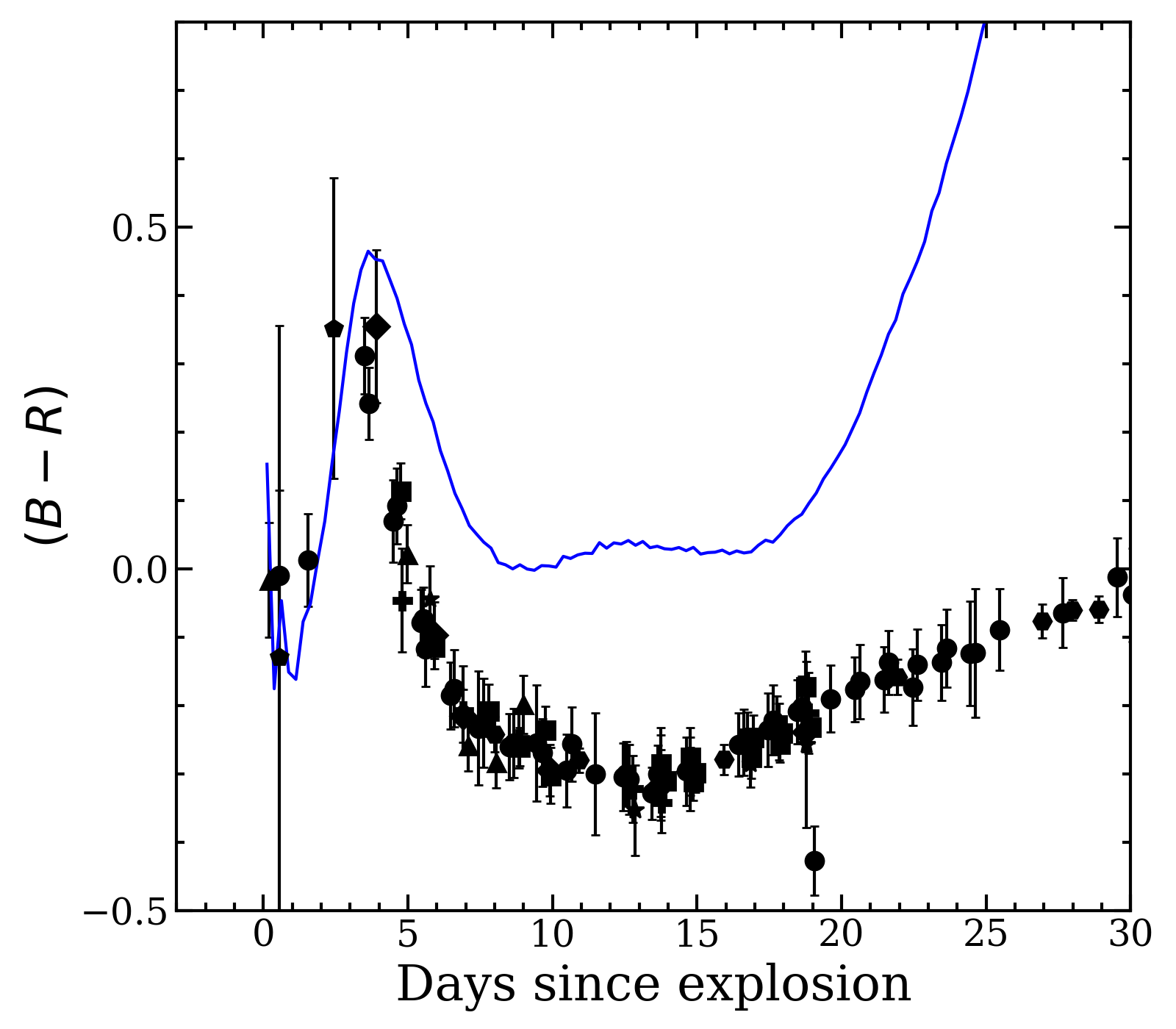}%
               \end{minipage}
               \caption{The light curve (Left), $B-V$ (Middle), and $B-R$ (Right) color evolution of SN 2021hpr with the He shell ($0.08\,M_{\odot}$) detonation on the $0.9\,M_{\odot}$ Sub-$M_{\rm ch}$ mass WD \citep{2019ApJ...873...84P}. The color and symbols are the same as in Figure 3. In the color curves, the model is presented as the blue solid line.}
            \end{figure*}

            \begin{figure*}\label{fig:figure12}
               \begin{minipage}{.32\textwidth}
                 \centering
                 \includegraphics[width=1.\textwidth]{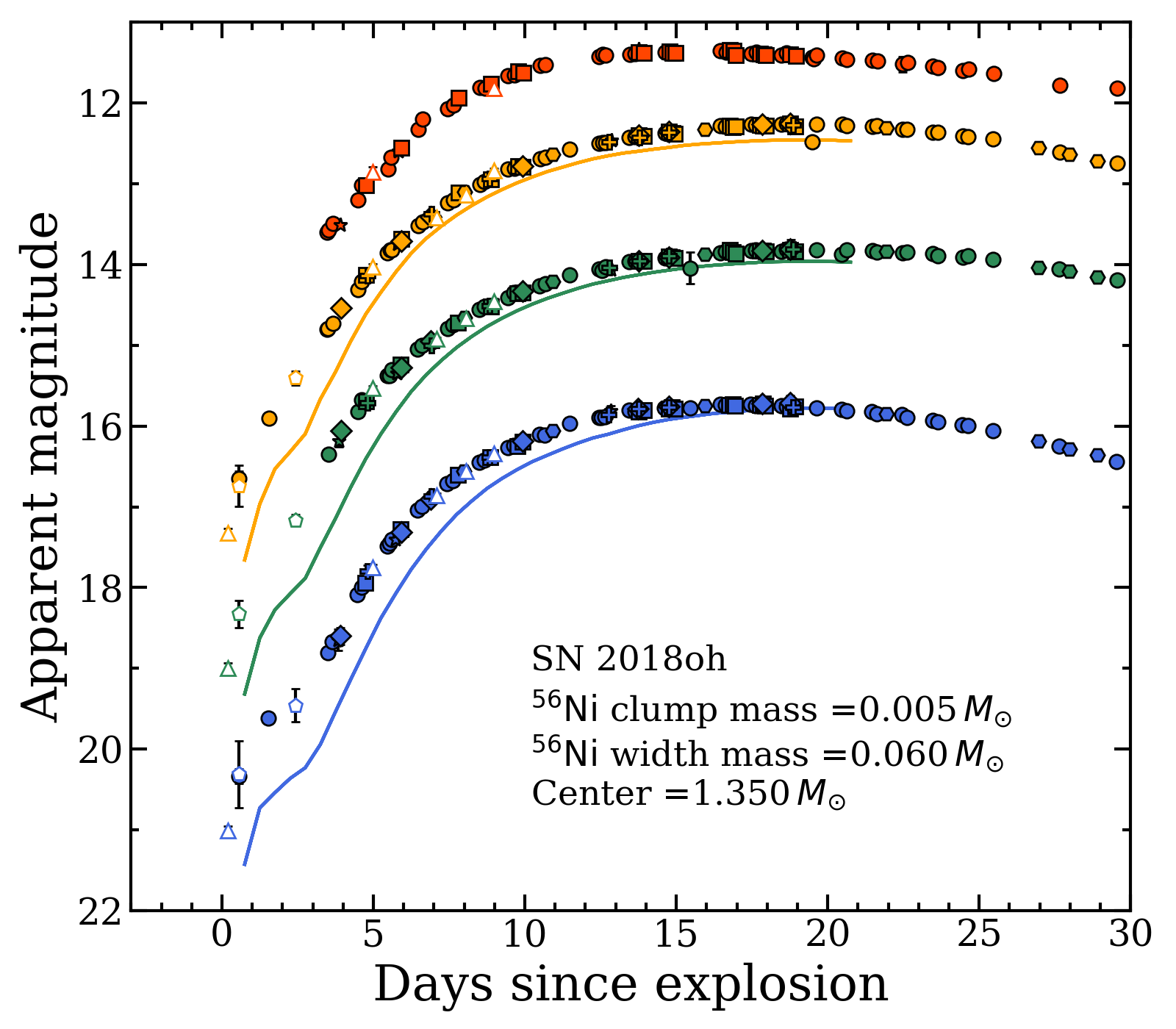}
               \end{minipage}
               \hfill
               \begin{minipage}{.32\textwidth}
                 \centering
                 \includegraphics[width=1.\textwidth]{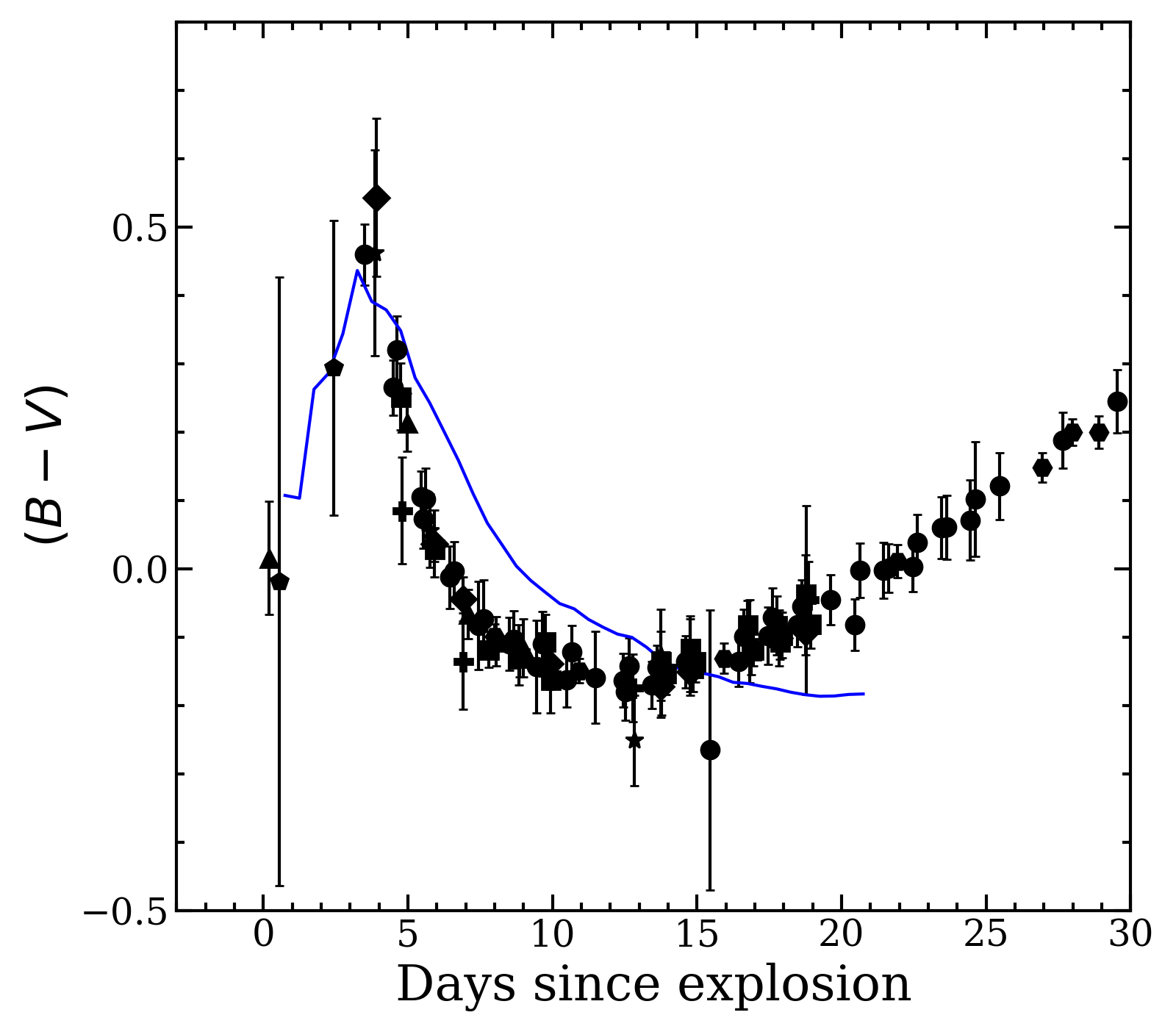}
               \end{minipage}
               \hfill
               \begin{minipage}{.32\textwidth}
                 \centering
                 \includegraphics[width=1.\textwidth]{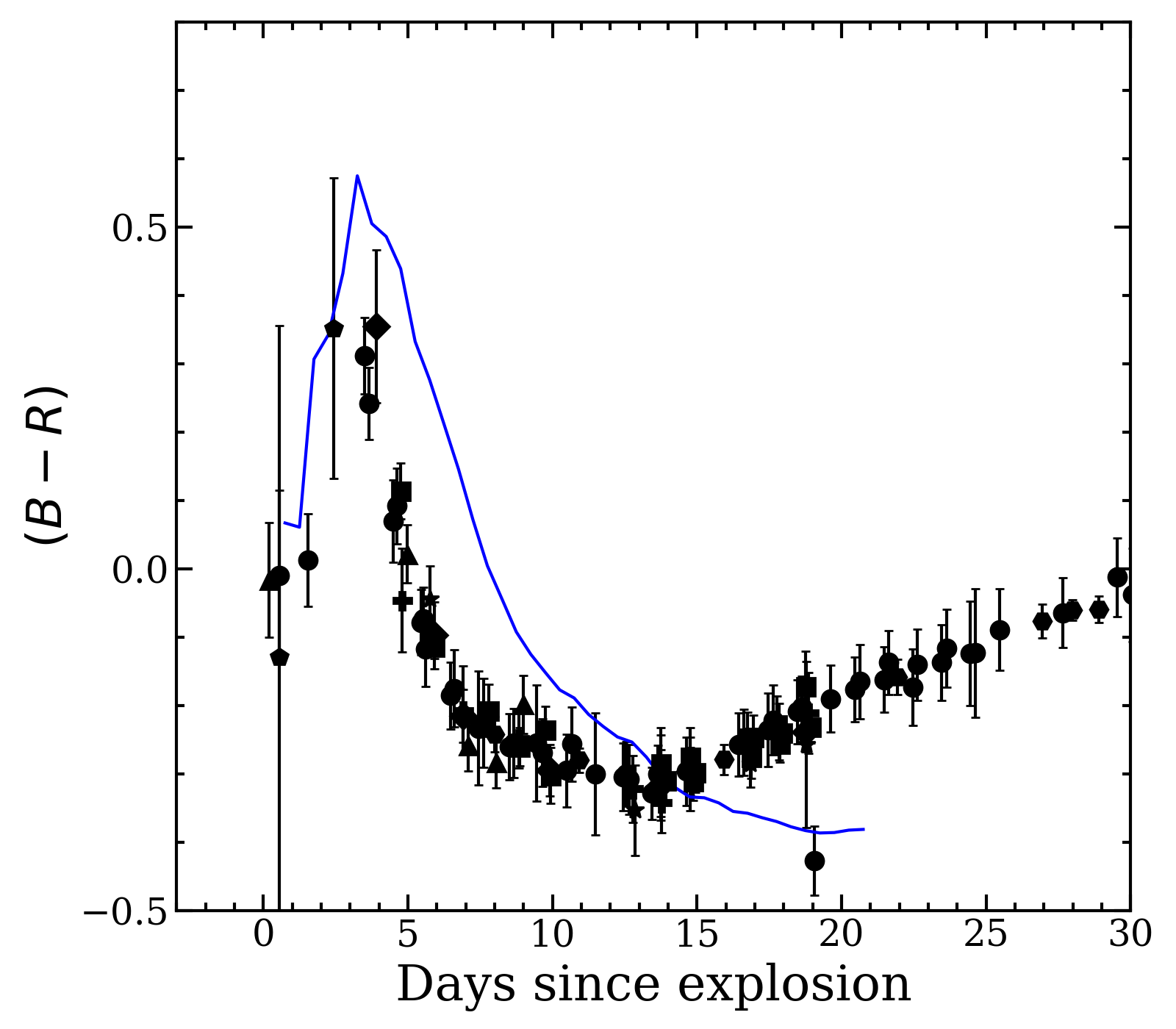}
               \end{minipage}
               \caption{Same as Figure 11 but with the $^{56}\rm Ni$ clump with $0.005\,M_{\odot}$ and with a width of $0.06\,M_{\odot}$ located on the mass coordinate of $1.35\,M_{\odot}$ based on the fiducial light curve of SN 2018oh \citep{2020A&A...642A.189M}.}
            \end{figure*}

            \begin{figure}\label{fig:figure13}
                \centering
                \includegraphics[width=1\linewidth]{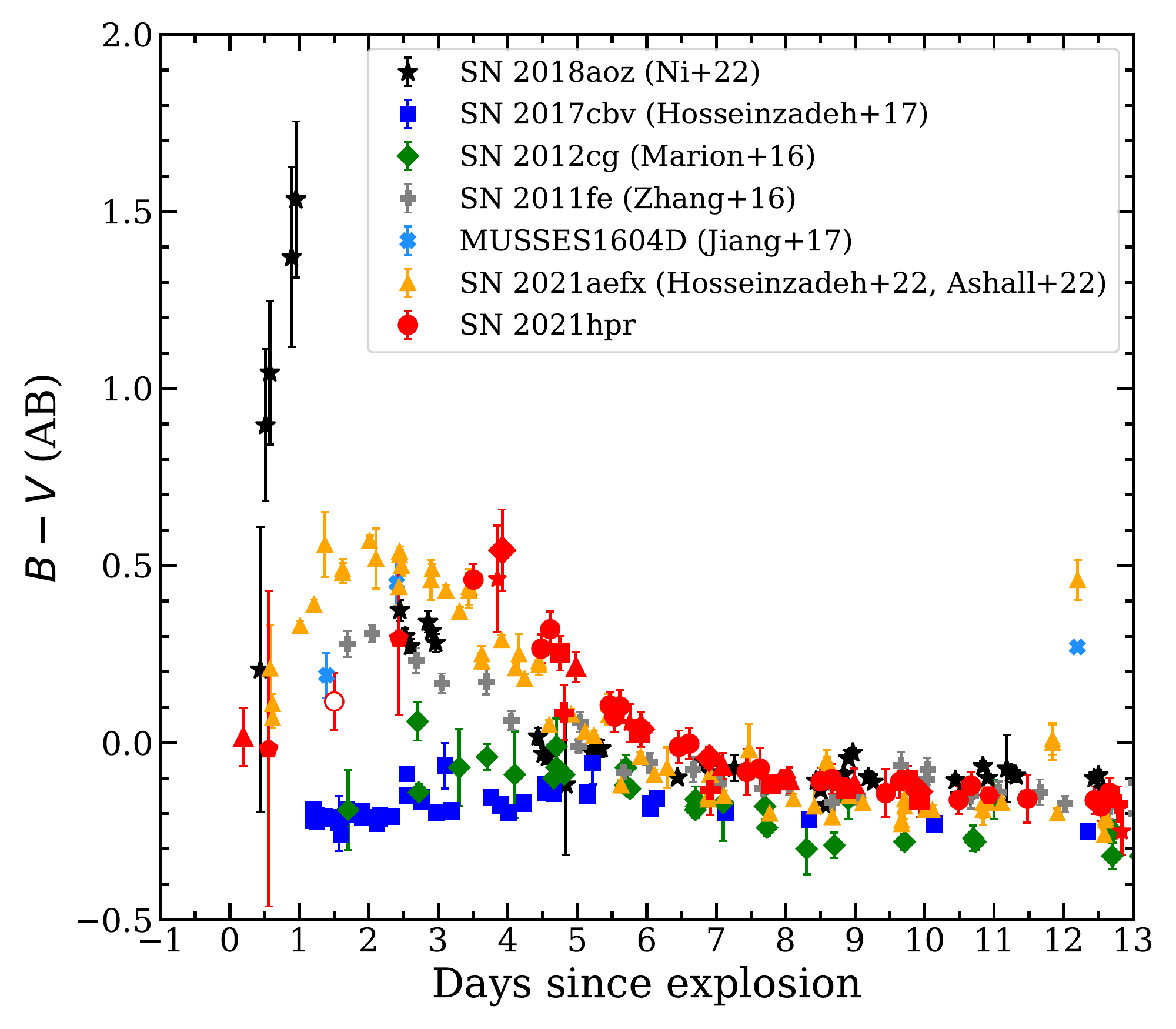}
                \caption{The de-reddened $B-V$ color evolution of SN 2021hpr with other type Ia supernovae within 13 days since the explosion. A data point for SN 2021hpr is added with a red open circle at $t \sim 1.5$ days by converting $B-R$ to $B-V$ using a correlation between the two quantities during the first 4 days since the explosion. For SN 2021aefx, we plotted both photometry in \citep{2022ApJ...932L...2A} including ground and $Swift$ data and \citep{2022ApJ...933L..45H}. Since there was no estimated explosion time in \citep{2022ApJ...933L..45H}, we use the explosion time ($59529.19$\,MJD) in \citep{2022ApJ...932L...2A}. The SNe Ia, except SN 2011fe, are known to have early excess. Symbols of SN 2021hpr are the same as Figure 3.}
            \end{figure}

    \subsection{Comparison of color curves with other SNe Ia}
        In Figure 13, we compare early color curves of several SNe Ia with an extensive set of early-time data, four with early excess (SN 2012cg, MUSSES1604D, SN 2017cbv, SN 2018aoz, and SN 2021aefx) and one without early excess (SN 2011fe). Figure 13 reveals a diversity of early color curves. The figure indicates that there are roughly four families of color curves, one with a blue, flat color curve (SN 2017cbv and SN 2012cg), one with a red peak at 2-3 days, and then either becoming redder again (MUSSES1604D) or blue (SN 2021hpr, SN 2011fe, and SN 2021aefx), and one that shows a red peak at a very early epoch ($\sim 1$\,day) and becomes blue (SN 2018aoz). These different behaviors may reflect the differences in the explosion mechanisms.

        For the first category of flat, blue curves of SN 2017cbv and SN 2012cg, \citet{2017ApJ...845L..11H} suggest a model with the circumstellar (CSM) material and nickel mixing. They disfavor the companion interaction model because such a model cannot explain the excess in Swift UV bands. It will also difficult to keep the color curve blue in the early epoch with a SD model which predicts an early red peak as in Figure 6.

        In the second category of SNe with an early red peak and a subsequent reddening, MUSSES1604D follows the DDet model trend well, where its red color can be explained by the presence of Fe-peak elements in the outer layer ejecta extinguishing blue light. The outer layer ejecta are possibly produced by He-shell in DDet models (\citealt{2017Natur.550...80J}).

        For the third category of SNe with an early red peak, followed by a blue light curve, SHCE can explain the observed properties as found for SN 2021hpr. Similarly, the $B-V$ color of SN 2021aefx reaches a peak ($\sim0.6$) at $\sim2$\, days, a bit faster than that of SN 2021hpr. Like SN 2021hpr, the early flux and color evolution of SN 2021aefx can be explained at least partly with the companion interaction \citep{2022ApJ...933L..45H}. However, there is no perfect explanation all of the observed properties of 
        SN 2021aefx for now despite many efforts in terms of the progenitor scenarios \citep{2022ApJ...933L..45H}.  On the other hand, the work of \citet{2022ApJ...932L...2A} cautions that the early UV emission excess of SN 2021aefx can be affected by the Doppler effect and the interpretation of the excess needs to take into account such effects. The lack of early excess emission for SN 2011fe may be due to a less optimal viewing angle suppressing the early excess emission or a very small companion star, although other possibilities (DDet models with thin He-shells) can be considered.

        For the last category of a very early red peak of SN 2018aoz, the companion shock-heating model is disfavored since such a model cannot reproduce the color curve behavior. The color behavior can be explained by an overabundance of Fe-peak elements due to burning in the extreme outer layer such as those in DDet models (\citealt{2022NatAs.tmp...45N}). Also, \citet{2022NatAs.tmp...45N} suggest a possibility of an extended subsonic mixing for the presence of outer layer Fe-peak elements.

        The light curves of SNe Ia can look alike and be parameterized as a uniform population, but a closer look of the early colors curves shows a wide variety of cases. This can be due to a variety of explosion mechanisms taking place. Hence, it is highly desired to expand the SNe Ia sample with very early multi-band light curves for a more statistical meaningful study.

\section{Summary} \label{sec:summary}
    We observed a Type Ia supernova, SN 2021hpr firstly reported on 2021 April 2.45 UT, using our IMSNG network of 0.4-1.0 meter class telescopes. A long-term light curve and a series of long-slit spectra show that SN 2021hpr is close to a normal type Ia supernova with a distance modulus of $33.28\pm0.11$ mag (d=$45.23\pm2.31\,$Mpc).

    Our analysis of early photometric data reveals distinct feature of SHCE. We fit the early data using a two-component model made of an ejecta-companion interaction component and a simple power-law component. The model explains the early excess in the light curve and color evolution. With an assumption of the optimal viewing angle, the best fit result is consistent with a companion radius of $8.84\pm0.58\,R_{\odot}$. The radius could be larger if the viewing angle is different. A subgiant star, a low mass red giant, or a helium star can be a possible donor but a low mass main sequence star is not likely to be the progenitor companion. We could not detect a probable progenitor candidate of SN 2021hpr in the HST archival deep pre-explosion images. The multi-band HST detection limits rule out massive stars with $M_{\rm init}>10\,M_{\odot}$ as the progenitor, giving us the upper limit on the radius of the progenitor system of $\sim215\,R_{\odot}$.

    Although the SD companion model can explain the early multi-band light curve evolution of SN 2021hpr, we could not find any strong signature of stripped mass ($\lesssim0.003\,M_{\odot}$ for H$\alpha$ emission) of H/He-rich material from the companion star in the late spectroscopy. This can result from a large binary separation rather than the low supernova explosion energy, but the non-detection of the nebular lines needs further theoretical and observational investigation. In particular, we did not analyze the early evolution of SN 2021hpr using DDet or DOM models in detail. Future, careful investigation may find that these two models can explain this distinct SN.

    To understand SNe Ia explosion mechanism, we compared color curves of SNe Ia with available very early $B-V$ data (available at $<1-2$\,days). These color curves have a diversity that can be summarized into four cases: (i) a color curve showing a very early red peak ($\lesssim$1\,day after explosion) like SN 2018aoz; (ii) color curves with a slower appearance of red peak ($2-3$\,days after explosion) with reddening of the curve in a later time (MUSSES1604D); (iii) or with the color staying blue after the red peak (SN 2021hpr, SN 2011fe, and SN 2021aefx); and (iv) flat, blue color curve (SN 2017cbv). The first two cases support the He-shell detonation or at least $^{56}\rm Ni$ in the outer ejecta, but a simple comparison of DDet models does not reproduce the early light and color curves of SN 2021hpr. A companion interaction model can explain the light and color curves of SN 2021hpr well, making this SN distinct from SNe Ia like SN 2018aoz (very early red peak), SN 2017cbv (flat blue color curve), and MUSSES1604D (slow red peak, late red color). Different early color properties of various SNe Ia suggest that the SN Ia explosion mechanism is diverse.

    The excellent agreement between the observed multi-band light curves and the SHCE model with a SD progenitor system is tantalizing but enigmatic with the no-detection of the nebular lines in the late-phase. An enlarged sample of SNe Ia with the data of this kind and an extensive study of the light curves should tell us how diverse the SNe Ia explosion mechanism can be. Also, the detection of early excess of SN 2021hpr demonstrates that high-cadence monitoring of nearby galaxies using small telescopes is a powerful tool to constrain the progenitor system of SN Ia even in the the era of large telescopes.

\acknowledgements
    This work was supported by the National Research Foundation of Korea (NRF) grants, No. 2020R1A2C3011091, and No. 2021M3F7A1084525 funded by the Korea government, and the Korea Astronomy and Space Science Institute under the R\&D program (Project No. 2022-1-860-03) supervised by the Ministry of Science and ICT (MSIT). G.L. acknowledges support from the Basic Science Research Program through NRF funded by MSIT (No. 2021R1A6A3A13045313) and MSIT (No. 2022R1A6A3A01085930). SCY is supported by the National Research Foundation of Korea (NRF) grant (NRF-2019R1A2C2010885). JCW and BPT are also supported by a DOE grant to the Wooten Center for Astrophysical Plasma Properties (WCAPP; PI Don Winget). JV is also supported by the project “Transient Astrophysical Objects” GINOP 2.3.2-15-2016-00033 of the National Research, Development and Innovation Office (NKFIH), Hungary, funded by the European Union. D.K. acknowledges support by the NRF of Korea (NRF) grant (No. 2021R1C1C1013580) funded by MSIT.

    The Hobby-Eberly Telescope (HET) is a joint project of the University of Texas at Austin, the Pennsylvania State University, Ludwig-Maximilians-Universität München, and Georg-August-Universität Göttingen. The HET is named in honor of its principal benefactors, William P. Hobby and Robert E. Eberly.

    The Low Resolution Spectrograph 2 (LRS2) was developed and funded by the University of Texas at Austin McDonald Observatory and Department of Astronomy and by Pennsylvania State University. We thank the Leibniz-Institut für Astrophysik Potsdam (AIP) and the Institut für Astrophysik Göttingen (IAG) for their contributions to the construction of the integral field units.

    We acknowledge the Texas Advanced Computing Center (TACC) at The University of Texas at Austin for providing high performance computing, visualization, and storage resources that have contributed to the results reported within this paper.

    We thank the staff at SAO, MAO, CBNUO, LOAO, SOAO, DOAO, and McDonald Observatories for their observations and maintenance of the facilities. This research made use of the data taken with LOAO and SOAO are operated by the Korea Astronomy and Space Science Institute (KASI), DOAO of National Youth Space Center (NYSC), HET of McDonald Observatory, and CBNUO of Chungbuk National University Observatory.


%\vspace{5mm}
\facilities{LOAO, SOAO, DOAO, MAO:$1.5\,$m, SAO, CBNUO:$0.6\,$m, McDonald Observatory:HET, HST}


\software{Python packages: Astropy \citep{Astropy13}, Numpy \citep{2020Natur.585..357H}, LMFIT \citep{2014zndo.....11813N}, PyRAF \citep{2012ascl.soft07011S}, Hotpants \citep{2015ascl.soft04004B}, Astronometry.net \citep{2010AJ....139.1782L}, SExtractor \citep{BA96}, AstroDrizzle \citep{2012drzp.book.....G}}




%% bibtext SN2021hpr.tex
%%pdflatex SN2021hpr.tex

%\begin{thebibliography}{9}
%\clearpage

{}

\appendix
\counterwithin{figure}{section}
\section{Figures}
\begin{figure*}
    \centering\offinterlineskip
    \includegraphics[width=1.1\linewidth]{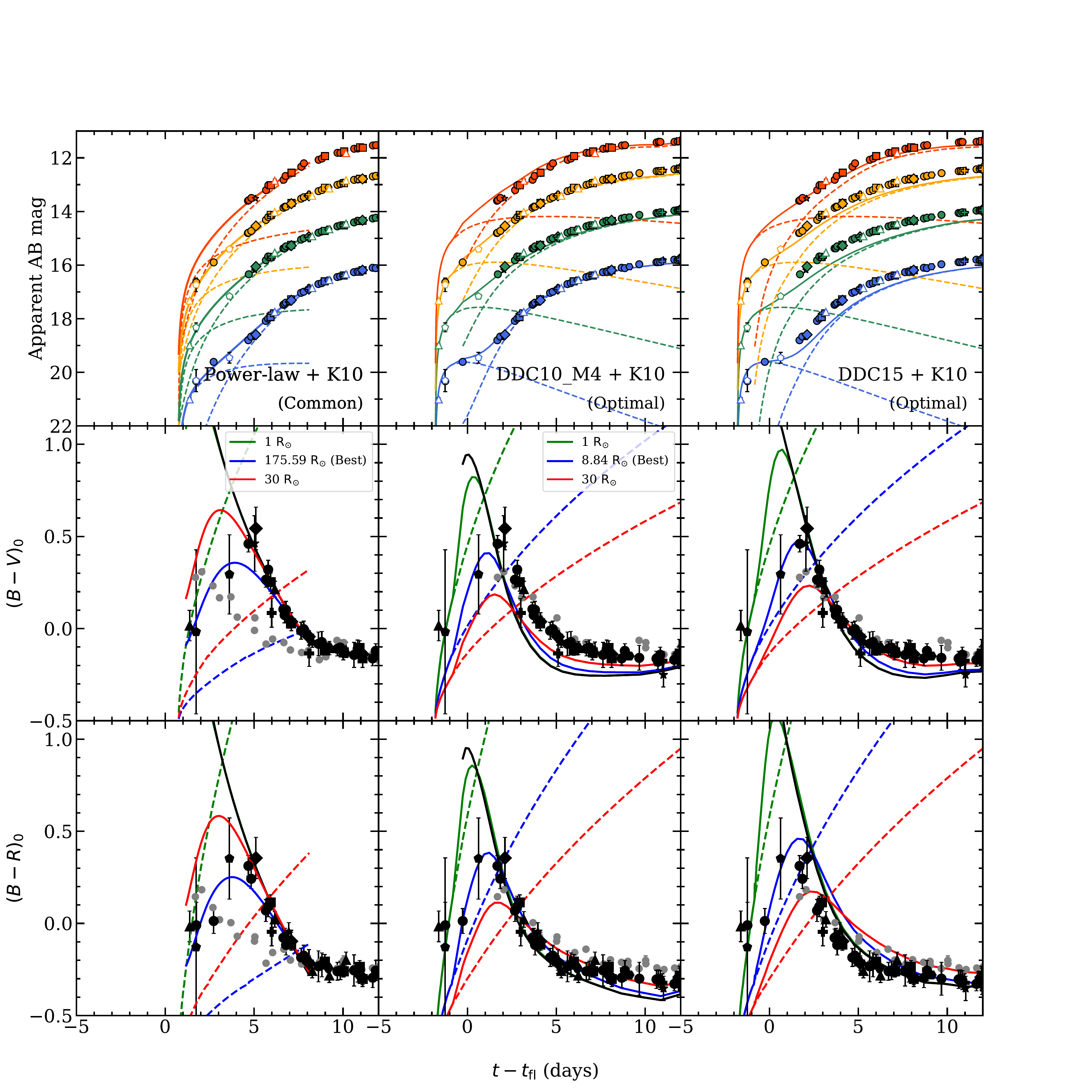}
    \caption{Same as Figure 6 with different models. (Left column) Companion+power-law fit, but note that it is for a common viewing angle. (Middle \& right columns) Cases for the optimal viewing angle using DDC10\_M4 (middle) and DDC15 model (right) with the best-fit SHCE model from Figure 6.}
\end{figure*}

\bibliographystyle{aasjournal}

\end{document}